
\input phyzzx

\def\bold#1{\setbox0=\hbox{$#1$}%
     \kern-.025em\copy0\kern-\wd0
     \kern.05em\copy0\kern-\wd0
     \kern-.025em\raise.0433em\box0 }

\def\epem{{e^+e^-}}
\def\eg{{\it e.g.}}

\def\M{{\cal M}}

\def\O{{\cal O}}
\def\brk{{\hfill\break}}
\def\gg{{\gamma \gamma}}
\def\tot{{\rm tot}}
\def\MS{{\overline{MS}}}
\pubnum{6571}
\date{July, 1994}
\pubtype{(T/E)}
\titlepage
\title{{\bf High Energy Photon--Photon Collisions}\doeack}
 \author{Stanley J. Brodsky}
\SLAC
\vskip .10in
\centerline{and}
 \author{Peter M. Zerwas}
\centerline{\it Deutsches Elektronen-Synchrotron DESY}
\centerline{\it D-22603 Hamburg FRG}
\vfill
\centerline{Presented at the  Workshop on}
\centerline{Gamma-Gamma Colliders}
\centerline{Lawrence Berkeley Laboratory, March 28-31, 1994}
\vfill
\endpage
\PHYSREV
\def\refout{\par\penalty-400\vskip\chapterskip
   \spacecheck\referenceminspace
   \ifreferenceopen \Closeout\referencewrite \referenceopenfalse \fi
   \line{\bf 21. References\hfil}\vskip\headskip
   \input \jobname.refs
   }
\hsize=6.0truein
\vsize=8.5truein
\singlespace
\overfullrule=0pt
\pagenumber=1
\hbox to \hsize{}
\vskip1cm
\sequentialequations
\Picture\figDD \height=.5in \width=\hsize
\caption{ \baselineskip=12pt\narrower
Factorization of the resolved photon-photon
amplitudes using the light-cone Fock
basis. (See Appendix I.)
In the case of the direct contributions, the photon annihilates within
the hard scattering amplitude.}
\savepicture\figDDpic

\Picture\Afiga \height=.5in \width=\hsize
\caption{ \baselineskip=12pt\narrower
The $\gamma \gamma$ luminosity in Compton
             back--scattering of laser light; unpolarized
$e^\pm$ beams and laser photons (dashed), opposite helicities
of $e^\pm$ and $\gamma$ (full curve).
See Refs. [\two], [\seven], [\KMS].}
\savepicture\Afigapic

\Picture\figH \height=.5in \width=\hsize
\caption{\singlespace
(a) Spectra,
(b) Stokes vector $\xi_3$ for linear polarization
from Ref. [\four],
and
(c) left/right asymmetry of the final state photon beam in
Compton back--scattering of laser light.
}
\savepicture\figHpic

\Picture\figBB \height=.5in \width=\hsize
\caption{  \baselineskip=12pt   \narrower
Illustration of direct, resolved,
and higher-order loop contributions to high
energy $\gg$ collisions.
}
\savepicture\figBBpic

\Picture\figAA \height=.5in \width=\hsize
\caption{ \baselineskip=12pt\narrower
Representative cross sections for $W^+ W^-$production
and other electroweak reactions at a $\gg$ and $e^+ e^-$ linear collider.
The top mass is taken as $130$ GeV. The other subscripts refer to the
mass of the Higgs (in GeV). The Higgs mass is set to zero
for the reactions $e^+ e^- \to W^+W^- \nu \bar \nu$ and
$e^+ e^- \to Z Z \nu \bar \nu.$
{}From Refs. [\JikiaWWWW,\BB,\BaillargeonRev].}
\savepicture\figAApic

\Picture\figY \height=.5in \width=\hsize
\caption{ \baselineskip=12pt\narrower
Differential cross sections for producing a $W$ pair of a specific
helicity
combination at $\sqrt s_\gg= 500$ GeV as a function of
$\cos\theta$.
The curves are:\brk
1: (++++)+(----), 2: (+++-)+(++-+)+(--+-)+(---+),\brk
3: (++--)+(--++), 4: (+-++)+(-+++)+(+---)+(-+--),\brk
5: (+-+-)+(-+-+)+(-++-)+(+--+),\brk
6: (+-+-)+(-+0-)+(+--0)+(-++0)+(+-+0)+(-+-0)+(+-0-)+(-+0+),\brk
7: (+-00)+(-+00), 8: (++0+)+(--0-)+(+++0)+(---0),\brk
9: (++00)+(--00), 10: (++0-)+(++-0)+(--0+)+(--+0).\brk
The notation indicates $(\lambda_1\lambda_2\lambda_3\lambda_4)$,
where $\lambda_1, \lambda_2, \lambda_3$ and $\lambda_4$ are the
helicities
of the two photons and the $W^+$ and the $W^-$ respectively.
{}From Ref.  [\Yehudai].~
}
\savepicture\figYpic

\Picture\figCC \height=.5in \width=\hsize
\caption{ \baselineskip=12pt\narrower
Standard Model one-loop contributions to the reaction $\gg \to Z Z$
including ghost $c^\pm$ and scalar $w$ contributions in the
background nonlinear gauge. From Ref.  [\JikiaNG].
}
\savepicture\figCCpic

\Picture\figEE \height=.5in \width=\hsize
\caption{ \baselineskip=12pt\narrower
The effective cross section for
$\gamma\gamma\rarrow Z^0 \gamma $  at an NLC taking into account
the back-scattered laser spectrum. The fermion and $W$ loop
contributions are shown for the production of a transversely (T) and
longitudinally (L) polarized $Z^0.$  The incident photons are taken to
have  positive helicity. From Ref.  [\JikiaNG].}
\savepicture\figEEpic

\Picture\Afigd \height=.5in \width=\hsize
\caption{ \baselineskip=12pt\narrower
Number of events per year for the Standard Model
Higgs boson $(\Phi^0 \rightarrow b \overline{b}, t \overline{t}, ZZ)$
and for the heavy--quark backgrounds; Ref. [\three].
Here ${\cal L}_{\rm eff}=20\, fb^{-1}$,
$z_0=0.85$, $\VEV{\lambda_1\lambda_2} = 0.8$,
$\Gamma_{\rm expt} = 5\ GeV$.}
\savepicture\Afigdpic

\Picture\Afige \height=.5in \width=\hsize
\caption{ \baselineskip=12pt\narrower
Expected event rates for the Higgs signal
and the background processes in $b \overline{b}, c \overline{c}$
two--jet final states for polarized $\gamma$ beams;
Ref. [\fourteen].}
\savepicture\Afigepic

\Picture\Afigf \height=.5in \width=\hsize
\caption{ \baselineskip=12pt\narrower
Invariant mass distribution in $\gamma \gamma
\rightarrow H \rightarrow ZZ$ and in the continuum  $ \gamma
\gamma \rightarrow ZZ$ for transverse and longitudinal $Z$
polarization; Ref. [\JikiaNG].}
\savepicture\Afigfpic

\Picture\Afigg \height=.5in \width=\hsize
\caption{ \baselineskip=12pt\narrower
The $\gamma \gamma$ polarization asymmetry ${\cal A}$
in Compton back--scattering of linearly polarized laser light
for various values of $x_0$; Ref. [\four].}
\savepicture\Afiggpic

\Picture\Afigh \height=.5in \width=\hsize
\caption{ \baselineskip=12pt\narrower
${\cal MSSM}$ Higgs particle $h^0$: signal and
background cross sections for $ b \overline{b}$ final states (a),
and the polarization asymmetry ${\cal A} (h^0)$ including the
background process (b); Ref. [\four].}
\savepicture\Afighpic

\Picture\Afigi \height=.5in \width=\hsize
\caption{ \baselineskip=12pt\narrower
${\cal MSSM}$ Higgs particle $A^0$: signal and
background cross sections for $ b \overline{b}$ final states (a),
and the polarization asymmetry ${\cal A} (A^0)$ including the
background process (b); Ref. [\four].}
\savepicture\Afigipic

\Picture\figEEE \height=.5in \width=\hsize
\caption{ \baselineskip=12pt\narrower
Illustration of $W W$ scattering at a photon-photon collider.
The kinematics of the
interacting $W$'s can be determined by tagging the spectator $W$'s.
The interacting pair can
scatter or annihilate, for example into a Higgs boson.}
\savepicture\figEEEpic

\Picture\figA \height=.5in \width=\hsize
\caption{\singlespace
(a) Deep--inelastic electron--photon scattering $e \gamma \to e X$.
(b) The charged current
charged current process $e \gamma \to \nu X$ in deep--inelastic
$e\gamma$ scattering.
}
\savepicture\figApic

\Picture\figB \height=.5in \width=\hsize
\caption{\singlespace
Event plane ${\cal P} = \left[\log Q^2,\log(1/x-1)\right]$ in
$e\gamma$ scattering ($Q^2$ is defined in GeV$^2$). Shown are
the two parallelograms which can be explored at LEP200 and LC500
and within which perturbative QCD can be applied.
}
\savepicture\figBpic

\Picture\figC \height=.5in \width=\hsize
\caption{\singlespace
QCD prediction for the photon structure $F_2^\gamma(x,Q^2)$ at
$Q^2 = 200$~GeV$^2$ and sensitivity to the QCD $\Lambda$
parameter. Error bars correspond to an integrated luminosity of
500~pb$^{-1}$ at LEP200 and the range $100 < Q^2 < 500$~GeV$^2$.
{}From Ref. [\Cordier].
}
\savepicture\figCpic

\Picture\figD \height=.5in \width=\hsize
\caption{\singlespace
Comparison of the $Q^2$ evolution of the photon structure
function in QCD with a model in which the coupling constant is
frozen.
}
\savepicture\figDpic

\Picture\figE \height=.5in \width=\hsize
\caption{\singlespace
Experimentally observed $Q^2$ evolution of the photon structure
function; from Ref. [\sasaki].
}
\savepicture\figEpic

\Picture\figI \height=.5in \width=\hsize
\caption{ \baselineskip=12pt\narrower
Comparison of  perturbative QCD predictions
with PLUTO data for the photon structure function at $Q^2=5.9$ GeV$^2.$
The charm quark contribution from leading and higher order
QCD  is also shown.  From Ref.  [\laenen].}
\savepicture\figIpic

\Picture\figF \height=.5in \width=\hsize
\caption{\singlespace
Theoretical estimate of the sensitivity to the effective
QCD scale
parameter, (a) from the evolution of $F_2^\gamma$ at large $x$;
(b) from the absolute size of $F_2^\gamma$ if the hadronic
component is assumed to be uncertain within $\pm 50\%$ at
$Q^2 = 100$~GeV$^2$.
}
\savepicture\figFpic

\Picture\figL \height=.5in \width=\hsize
\caption{ \baselineskip=12pt\narrower
QCD contributions to jet transverse momentum
cross section in $\gg$ collisions.
The resolved contributions are based on
the Drees-Godpole model for the gluon distribution in the photon. From
Ref. [\Resolved]. }
\savepicture\figLpic

\Picture\figO \height=.5in \width=\hsize
\caption{ \baselineskip=12pt\narrower
The effect of multiple scattering on the
mini-jet contribution to the $\gg$
total cross section. The jet contributions are shown for various
$p_T$ minimum cut-offs, with (solid line) and without
(dashed line) the effect of eikonalization. From
Ref.  [\FS].}
\savepicture\figOpic

\Picture\figWW \height=.5in \width=\hsize
\caption{ \baselineskip=12pt\narrower
Perturbative QCD contributions to large momentum transfer
exclusive double diffractive $\gg$ processes. The two-gluon
exchange pomeron  contributions to vector meson pair production and
three-gluon exchange odderon contributions to neutral pseudoscalar and
tensor meson pair production are illustrated.}
\savepicture\figWWpic

\Picture\figR \height=.5in \width=\hsize
\caption{ \baselineskip=12pt\narrower
Direct and resolved contributions to heavy quark production in $\gg$
collisions.
}
\savepicture\figRpic

\Picture\figS \height=.5in \width=\hsize
\caption{ \baselineskip=12pt\narrower
QCD leading and next-to-leading order contributions
to the inclusive charm
production cross section. The resolved contributions are based on
the Drees-Godpole model for the gluon distribution in the photon. From
Ref. [\DKZZ].
}
\savepicture\figSpic

\Picture\figT \height=.5in \width=\hsize
\caption{ \baselineskip=12pt\narrower
TPC$\gg,$  TASSO, and TRISTAN data for $\sigma(\epem\rarrow \epem
D^{\ast\pm}X) $ compared with the QCD prediction of Drees \etal\
{}From Ref.  [\DKZZ].
The dashed lines show the once-resolved contributions.
The upper line is $\mu = m_c = 1.3$ GeV.  The lower line is
$\mu = 2m_c,\ m_c = 1.8$ GeV.
}
\savepicture\figTpic

\Picture\figU \height=.5in \width=\hsize
\caption{ \baselineskip=12pt\narrower
(a) Effective cross section $\VEV{\sigma(\gg \to t {\bar t})}/
\sigma(\epem)_{pt}$
with (solid) and without (dashed) QCD corrections for $m_t=150~GeV.$ The
convolution with a back-scattered laser spectrum $(\omega=1.26$ eV) is
included.
(b) The  effective differential cross section
${d\VEV{\sigma(\gg \to t {\bar t})}/dz / \sigma(\epem)_{pt}}$
and the resonance
signal
predicted at $\sqrt{s_\epem} = 500$ GeV. Here $z = \M_{t \bar
t}/\sqrt{s_\epem}.$  From Ref.  [\KMS]. }
\savepicture\figUpic

\Picture\figQQ\height=.5in \width=\hsize
\caption{ \baselineskip=12pt\narrower
The single top production cross section
and its competing
backgrounds in high energy electron-photon collisions.
{}From Ref. [\Cheung].}
\savepicture\figQQpic

\baselineskip=17pt

\vglue 1cm
\centerline{\fourteenbf
HIGH ENERGY PHOTON--PHOTON COLLISIONS$^*$}
\vglue 1cm
\centerline{\rm STANLEY J. BRODSKY}
\baselineskip=15pt
\centerline{\it Stanford Linear Accelerator Center}
\centerline{\it Stanford University, Stanford, California 94309}
\centerline{\it e-mail: sjbth@slac.stanford.edu ---
                       phone: (415) 926-2644 ---
                         fax: (415) 926-2525}
\vskip .10in
\centerline{and}
\vskip .10in
\centerline{\rm PETER M. ZERWAS}
\centerline{\it Deutsches Elektronen-Synchrotron DESY}
\centerline{\it D-22603 Hamburg FR Germany}
\centerline{\it e-mail: T00ZER@DHHDESY3  ---
                       phone: +49.40.8998.2416 ---
                         fax: +49.40.8998.2777 }

    \vfootnote*{Work supported in part
by the Department of Energy,
      contract DE--AC03--76SF00515.}

\vskip 1cm
\centerline{\tenrm ABSTRACT}
The collisions of high energy photons
produced at an electron-positron collider
provide a comprehensive laboratory for testing QCD, electroweak
interactions, and extensions of the standard model.
The luminosity and energy of the colliding photons produced by
back--scattering  laser beams is expected to be comparable
to that of the primary $\epem$ collisions.  In this
overview, we shall focus on tests of electroweak theory in photon-photon
annihilation, particularly $\gg \to W^+ W^-$,
$\gg \to $ Higgs bosons, and
higher-order loop processes, such as $\gg \to \gg, Z \gamma$ and $Z Z.$
Since each photon can be resolved into a $W^+ W^-$ pair, high energy
photon-photon collisions can also provide a remarkably background-free
laboratory for studying $W W$ collisions and annihilation.
We also review
high energy $\gg$ tests of quantum chromodynamics, such as the scaling of
the photon structure function, $t \bar t$ production, mini-jet processes,
and diffractive reactions.
\vglue 0.5cm

\endpage
\baselineskip 22pt

\vskip1cm\leftline{{\bf 1. Introduction}}
\vskip.5cm

\REF\AppA{%
The Fock state decomposition of the photon
and a description of photon-photon collision processes at
the amplitude level are discussed in Appendix I.}

\REF\Ioffe{B. L. Ioffe, \sl Phys. Lett. \bf 30B \rm (1969)
123; J. Pestieau, P. Roy, and H. Terazawa, \sl Phys. Rev. Lett.
\bf 25 \rm (1970) 402; A. Suri and D. R. Yennie, \sl Ann.
Phys. (N.Y.) \bf 72 \rm (1972) 243.
See also,
L. Stodolsky, in the
{\it Proceedings of the
International School of Elementary
Particle Physics}, Herceg-Novi,
Yugoslavia, (1969).
            Edited by M. Nikolic.
        N.Y., Gordon and Breach, 1977.}

\REF\Yennie{%
T. H. Bauer, R. D. Spital, and D. R. Yennie,
\sl Rev. Mod. Phys. \bf 50 \rm (1968) 261.}

\REF\AppB{%
For further discussion of the photon formation time, see
Appendix II.}

\REF\SchulerA{%
For a recent review of the hadronic aspects of
photon-photon collisions,
see G. A.  Schuler and T.  Sj\"ostrand, CERN-TH-7193-04,
presented at Two-Photon Physics from DAPHNE to LEP200 and Beyond,
Paris, France, 2-4 Feb 1994.
(1994).}

\REF\BKT{%
S. J. Brodsky, T. Kinoshita, and H. Terazawa,
\sl Phys. Rev. Lett. \bf 25 \rm (1970) 972;
\sl  Phys. Rev. \bf D4 \rm (1971) 1532.}

\REF\Kessler{%
For a review of additional aspects of
hadronic $\gg$ physics and further references,
see P. Kessler, {\it Proceedings of the IXth International Workshop on
Photon-Photon Collisions}, D. Caldwell
and H. Paar, eds. (World Scientific, 1992.)}

\REF\walsh{%
T. F. Walsh,
\sl Phys. Lett. \bf 36B \rm (1971) 121.}

\REF\Witten{
E. Witten, \sl Nucl. Phys. \bf B120 \rm (1977) 189.}

\REF\LepageBrodsky{%
G. P. Lepage and S. J. Brodsky,
\sl Phys. Rev. \bf D22 \rm (1980) 2157.
S. J. Brodsky and G. P. Lepage,
\sl Phys. Rev.  \bf D24 \rm (1981) 1808;
                \bf D24 \rm (1981) 2848.}
\REF\Resolved{%
S. J. Brodsky, T. DeGrand, J. Gunion, and J. Weis,
\sl Phys. Rev. Lett. \bf 41 \rm (1978)  672;
\sl Phys. Rev. \bf 19, \rm (1979) 1418.
The  descriptive terms,
direct and resolved, were introduced by
M. Drees and R. M. Godpole, these proceedings, \sl Nucl. Phys.
\bf B339 \rm (1990)  355;
\sl Phys. Rev. Lett. \bf 67 \rm (1992) 1189;
and  DESY 92-044 (1992). M. Drees, DESY 92-065
(1992).}

The photon, as postulated nearly a century ago by Planck,
is a   discrete quantum of  electromagnetic energy.
The advent of TeV electron-positron
linear colliders will allow the study of
the collisions of beams of
photons with  energies a trillion times higher than
those of ordinary light.

In quantum field theory,  the electromagnetic field  couples to all
particles carrying the electromagnetic current,
and  thus a photon can fluctuate
into virtual states of remarkable complexity\refmark\AppA.
At high energies, the fluctuation of a photon
into a Fock state of particles of total
invariant mass $\M$ can persist over a
time of order $\tau = 2 E_\gamma /\M^2$  --
until the virtual state is materialized by a
collision or annihilation  with
another system\refmark{\Ioffe--\AppB}.
For example, in quantum chromodynamics,
the photon will couple through each quark
current into a  spectrum of  virtual
meson-like  color-singlet charge-zero
hadronic states.  The cross section for the production of
hadrons in the high energy collision of two photon beams will
thus resemble
the cross section for the collision of  ensembles of
high energy mesons\refmark{\SchulerA--\Kessler}.
In  the case of $e \gamma$ collisions,
the electron can scatter on the quark Fock states
of the photon, and one can study the
shape and evolution of  both
unpolarized and polarized photon structure
functions\refmark{\walsh,\Witten}\
$F_i^\gamma(x,Q^2),
g_i^\gamma(x,Q^2)$ in close analogy  to the
study of proton structure functions in  deep inelastic
lepton-nucleon
scattering. In  electron-photon collisions where the final state
is completely determined,  such as $\gamma e \to e
M^0,$ one can measure  photon--to--meson transition form
factors and other exclusive channels in analogy to the
transition form factors measured in exclusive electron proton
processes\refmark\LepageBrodsky.

\REF\SJB{For a review of two-photon exclusive processes, see
S. J. Brodsky,
{\it Proceedings of the IXth International
Workshop on Photon-Photon Collisions,}
D. Caldwell and H. Paar, eds. (World Scientific, 1992.)}

\REF\Hyer{%
T. Hyer, \sl Phys. Rev. \bf D47 \rm (1993) 3875.}

\REF\Ginzburg{%
See I. F. Ginzburg,
{\it Proceedings of the Second International Workshop
on Physics and Experiments with Linear Colliders,}
Waikoloa, Hawaii (1993),
Ed. F.~A.~Harris \etal, World Scientific.
and  the {\it Proceedings of the IXth
International Workshop on Photon-Photon Collisions,}
D. Caldwell and H. Paar,
eds. (World Scientific, 1992.)}

Thus two-photon collisions
can provide an important laboratory for
testing many types of coherent and incoherent
effects in quantum chromodynamics.
In events where each photon is resolved\refmark\Resolved~ in terms of
its intermediate quark and gluon states,
high momentum transfer photon-photon
collisions resemble hard meson-meson collisions
as illustrated in Fig. \figDD.
In the case of
exclusive final states such as $\gg \to p \bar p$ or meson pairs,
photon-photon collisions  provide a  time-like Compton microscope for
measuring  distribution amplitudes, the fundamental wavefunction
of hadrons\refmark{\SJB,\Hyer}.
One can study detailed features of $\gg \to t \bar t$ at threshold
and its final
state evolution. In the case of single or double diffractive
two-photon events,
one can study fundamental aspects of pomeron and odderon
$t$-channel physics\refmark\Ginzburg.

\REF\Ohnemus{%
The production of non-strongly interacting supersymmetric particles
in $\gg$ collisions is discussed by J. Ohnemus, T. F. Walsh,
and P.  M.  Zerwas, DESY 93-173 (1993).}

\REF\KolanoskiZerwas{%
H. Kolanoski and P. Zerwas, DESY 87-185 (1987),  published
in ``High Energy Electron-Positron Physics,", eds.
A. Ali and P. S\"oding, World Scientific, Singapore. }

\REF\Field{%
J. H. Field, preprint L.P.N.H.E. 84-04 (1984).}

\REF\Wagner{%
Ch. Berger and W. Wagner, \sl Phys. Rep. \bf 136 \rm (1987) 1.}

\REF\Kolanoski{%
H. Kolanoski, Two-Photon Physics at $\epem$ Storage Rings,
Springer Verlag,
1984)}

\REF\SJBshoresh{%
S. J. Brodsky, in the
{\it Proceedings of the IXth International Workshop on
Photon-Photon Collisions}, Shoresh, Israel (1988).}

\REF\BBCRev{%
D. L. Borden, D. A, Bauer, D. O. Caldwell, SLAC-PUB-5715 (1992).}

\REF\BaillargeonRev{%
For a comprehensive review of electroweak physics at a high energy
$\gg$ collider, see M. Baillargeon,
G. Belanger, and F. Boudjema, ENSLAPP-A-473 (1994).}

\REF\BBW{%
For a review of the $\gg \to W W$  process,
and further references see M. Baillargeon and F.
Boudjema,
\sl Phys. Lett. \bf B288 \rm (1992) 210.}

\REF\Yehudai{%
E. Yehudai, {\it Proceedings of the Second International Workshop
on Physics and Experiments with Linear Colliders},
Waikoloa, Hawaii, (1993),
Ed. F.~A.~Harris \etal, World Scientific;
SLAC-383 (1991);
\sl Phys. Rev. \bf D44 \rm (1991) 3434;
\sl Phys. Rev. \bf D41 \rm (1990) 33.}

\REF\BrodskyW{%
S. J. Brodsky, {\it Proceedings of the Second International Workshop
on Physics and Experiments with Linear Colliders},
Waikoloa, Hawaii,(1993),
Ed. F. A. Harris \etal, World Scientific, Vol I, p. 295. }

\REF\BrodskyJikia{%
S. J. Brodsky,  and G. Jikia, in progress.}

\REF\JikiaWWWW{%
G. Jikia ,  these proceedings;  and Protovino preprint (1994),
hep-ph/9406395. }

\REF\Cheung{%
K. Cheung, {\it Proceedings of the Second International Workshop
on Physics and Experiments with Linear Colliders},
Waikoloa, Hawaii (1993),
Ed. F. A. Harris \etal, World Scientific.
Northwestern preprint   NUHEP-TH-93-3, (1993).
See also K. Cheung, \sl Phys. Rev. \bf D47 \rm (1993) 3750.
D. Bowser-Chao, K. Cheung, and S. Thomas,
Northwestern preprint NUHEP-TH-93-7 (1993).}

\REF\CheungWWWW{%
K. Cheung, \sl Phys. Lett. {\bf B323} \rm (1994) 85,
Northwestern preprint  NUHEP-TH-94-13 (1994), and these proceedings.}

In addition to quarks, leptons, and $W's$, the photon couples to  all
particle-antiparticle pairs
postulated to  carry the electromagnetic current:
charged Higgs, supersymmetric
particles\refmark\Ohnemus, etc.
Thus high energy $\gg $ collisions can provide a laboratory for
exploring virtually every aspect of the  Standard Model and its
extensions\refmark{\KolanoskiZerwas--%
\BaillargeonRev}.   Two
photons can directly annihilate into $W$ pairs\refmark{\BBW,\Yehudai}~
or $q \bar
q$ pairs at the tree graph level, or pairs of gluons, pairs of photons,
$Z^0$'s, or one or more Higgs
bosons through quark and $W$ box graphs.  Two real photons can  couple to
any even charge conjugation resonance, unless it has spin  $J=1,$
in which case it can be identified via its virtual photon couplings.
In $\gg$ events where each  incident
photon produces a $W$-pair, the
$\epem$ collider becomes the equivalent of
a tagged $W W$ collider, allowing the study of $W W$ scattering and
annihilation in a  new domain of electroweak and Higgs
physics\refmark{\BrodskyW--\CheungWWWW}.

\REF\Telnov{%
I. Ginzburg, G. Kotkin, V. Serbo, and V.~Telnov,
\sl JETP Lett. \bf 34 \rm (1982) 491. For further references and
reviews see the contributions of V.
Telnov, V. Balakin and I. F. Ginzburg, and D.
Borden, {\it Proceedings of the Second International Workshop
on Physics and Experiments with Linear Colliders},
Waikoloa, Hawaii, (1993), Ed. F. A. Harris \etal, World Scientific,
and the reports of V. Telnov and J. Spencer in the
{\it Proceedings of the IXth International Workshop
on Photon-Photon Collisions}, D.
Caldwell and H. Paar, eds. (World Scientific, 1992.)}

All of the physics programs which we will discuss in this report appear
to be experimentally feasible at a high energy linear $e^+e^-$ collider,
since by using back-scattered laser
beams  (see Section 2),
it is expected that  the  $\gg$ luminosity will be
comparable to the electron-positron luminosity, and that  high
photon energy  and polarization can be attained\refmark\Telnov.
Thus it is clear that a central focus of investigation
at  the next  electron-positron linear collider will be the study of
photon-photon and electron-photon collisions.
\endpage

\vskip1cm\leftline{{\bf 2. Sources of $\bold{\gg}$ Collisions}}
\vskip.5cm

In an $e^+ e^-$ or $e^- e^-$  linear collider
there are three main sources of photon-photon
collisions.
The first is the equivalent photon spectrum
in which the virtual Weizs\"acker--Williams bremmstrahlung has
the relatively soft spectrum
$
G_{\gamma/e}(x,Q^2)\sim {\alpha\over 2\pi}\ \log\, {s\over M^2_e}\
{1+(1-x)^2\over x}.$  The equivalent photon approximation applied
to each of the incident leptons gives cross sections
analogous to the QCD factorization formula for fusion processes in high
transverse momentum inclusive reactions\refmark\BKT.
Virtual bremmstrahlung has been
the traditional mode for studying two-photon physics
at $\epem$ storage rings, and it will continue to
be very important at the next generation of B-factories.
By tagging the
scattered electron one can also select photons with a given
spacelike mass and polarization.
Thus, in the case of tagged electron-electron or electron-photon
collisions, the photon mass itself becomes a variable.

\REF\BCK{%
The spectrum of beamstrahlung photons depends on the machine design; see
T. Barklow,  P. Chen, and  W. Kozanecki,  SLAC-PUB 5718 (1992) and
Proceedings  ``$e^+e^-$ Collisions at 500~GeV: The Physics Potential'',
                  Munich--Annecy--Hamburg 1991, DESY 92-123B.}

\REF\DrellBlankenbecler{%
R. Blankenbecler and S. D. Drell,
\sl Phys. Rev. Lett. \bf 61 \rm (1988) 2324;
\sl Phys. Rev. \bf D37 \rm (1988) 3308;
\sl Phys. Rev. \bf D36 \rm (1987) 277.}

\REF\WuJacob{%
M. Jacob and T. T. Wu, in the {\it Proceedings of the
Lepton/Photon Symposium}  (1991);
\sl Phys. Lett. \bf B216 \rm (1989) 442;
\sl Phys. Lett. \bf 197B \rm (1987) 253.}

The low repetition rate of $e^+e^-$
linear colliders requires very small transverse dimensions of the
electron positron bunches in order to provide the necessary collision
rate. As a result, the particle trajectories are bent by the strong
electromagnetic fields within the bunches, giving rise to the
emission of synchrotron light\refmark{\BCK}.
Depending on the geometric shape of the bunches, hard real $\gamma$
spectra are generated, in particular for beams with small
transverse aspect ratio.
Thus beams of real photons will be automatically
created in high luminosity linear colliders by  the
``beamstrahlung"  process in which
an electron going through the opposing  high density bunch of
positrons
scatters and radiates a  spectrum of  nearly collinear
photons\refmark{\DrellBlankenbecler,\WuJacob}.
The beamstrahlung  photons are
unpolarized, but the spectrum  can be considerably
harder than that of the corresponding equivalent
photon spectrum\refmark\BCK.

It appears that the
most advantageous way to initiate photon-photon collisions at the
next linear collider will be to use a back-scattered laser beam, as
pioneered by Ginzburg \etal\refmark\Telnov~ In this process, photons
from a laser beam of eV energies are scattered against an electron
or positron beam to produce a nearly collinear beam of high energy
photons.

If the laser photons and the incoming $e^\pm$ beams are unpolarized,
the $\gamma \gamma$ luminosity of the Compton back--scattered
high--energy photons depends only weakly on the invariant
energy. However, the spectrum can be made hard by scattering
circularly polarized laser photons on polarized electrons/positrons
of opposite helicity as shown in  Fig.~\Afiga.
The peak of the spectrum is close
to the maximum of the $\gamma \gamma$ energy at $y_{\max} =
x_0/(1 + x_0)$ where $x_0 = 4E_e \omega_0/m^2_e$ is the invariant
$(\gamma e)$ energy--squared in units of the electron mass.
To avoid non--controllable $e^+e^-$ pair production in collisions
of laser $\gamma$'s with high energy $\gamma$'s, $x_0$ must
be chosen less than 4.83. Near this limit, about 80\% of the
$e^+e^-$ energy is transferred to the $\gamma \gamma$ system.
At the same time, the high energy photons are circularly
polarized at a degree of nearly 100\% in the peak region
[see Fig.~\figH (a)].

The transfer of the linear polarization from the laser
photons to the high--energy photons  is  also highly efficient.
The degree of the linear polarization of the high--energy
$\gamma$ beam is described by the third component of the
Stokes vector which reaches the maximum value at the upper
limit of the energy transfer, $\xi^{\max}_3 =  2 \,(1 + x_0) /
[1 + (1 + x_0)^2]$ [see Fig.~\figH (b)]. The degree of linear
polarization rises if $x_0$ decreases; for $x_0 = 1$ we
obtain $\xi^{\max}_3 = 4/5$, {\it i.e.} 80\%\ polarization
in the high-energy photon beam.

Unlike the beamstrahlung and equivalent photon processes, the
effective $e^+ e^-$ and  laser-induced high energy $\gamma
\gamma$ luminosities can be comparable.  The extraordinary energy
and high luminosity
of the back-scattered laser collisions promise to make two-photon
physics a key
component of the physics program of the next linear collider.

\REF\MillerA{%
D. Miller, {\it Proceedings of the Second International Workshop
on Physics and Experiments with Linear Colliders},
Waikoloa, Hawaii, (1993),
Ed. F. A. Harris \etal, World Scientific.}

In principle, each of the three types of photon beams can collide with
each other, so there are actually nine  possible $\gamma \gamma$
collisions at a linear collider;  one also has the  possibility of real
photons colliding with tagged  virtual photons through photon-
electron collisions\refmark\MillerA.
\endpage

\vskip1cm\leftline{{\bf 3. Survey of Photon-Photon Collider Processes}}
\vskip.5cm

\REF\Jikiagg{%
G. Jikia and A. Tkabladze,
\sl Phys. Lett. \bf B323 \rm (1994) 453.}

\REF\DSZ{A. Djouadi, M. Spira, and P. M. Zerwas, DESY 92-170.
K. Melnikov and O.~Yakovlev, Novosibirsk BUDKERINP preprint 93-4 (1993).}

Figure \figBB\ illustrates many of the
processes which could be studied at a high energy $\gg$ collider.
\pointbegin
The simplest reactions are the
direct $\gamma\gamma$ couplings  to pairs of leptons, $W$'s,  and
quarks.   Any energetically accessible particle which carries
the electromagnetic charge, including supersymmetric  and  technicolor
particles, can be produced in pairs.  In each
case, the charged line  can then radiate
its respective gauge partners: \eg, photons, gluons, $Z$'s  as well as
Higgs bosons.
\point
As shown in Fig. \figBB b, one can produce pairs of
charge-less fundamental particles in $\gamma \gamma$ collisions
through quantum loop diagrams, as in the traditional light-by-light
scattering box graph\refmark\Jikiagg.
For example, two photons can annihilate and produce two outgoing
photons or a pair of co-planar
$Z$'s through virtual  $W$ and quark loops. A pair of gluon
jets  can be produced through a  quark box diagram.  A single Higgs
boson or an excited $Z'$
can be produced through triangle graphs\refmark\DSZ.
\point
At high energies, one or both of the
an incident photons can ``resolve" itself as a pair of
fundamental charged particles
which can then interact  via scattering
subprocesses.
Thus, as illustrated in Fig. \figBB c, a  photon can develop
into a Fock state of $q\bar q$~ or $W^+W^-$~ or leptonic
pairs,   which then interact by $2 \to 2$ processes;  \eg,  quark-quark
scattering through gluon exchange
or top-quark scattering through Higgs exchange\refmark\AppA.
In addition, one can have interactions of a directly coupled photon
with the  resolved constituent of the other photon.

The cross sections for a number of  electroweak
processes that can be studied in a high energy linear
collider are shown in Fig. \figAA.
\endpage

\vskip1cm\leftline{{\bf 4. $\bold{\gg} \to W^+ W^-$ Production
in a $\bold{\gg}$ Collider}}
\vskip.5cm

One of the most important applications of two photon physics is the
direct production of $W$ pairs. By using polarized back-scattered
laser beams, one can in principle study $\gg \to W^+ W^-$
production as a function of the initial
photon helicities as well as resolve the $W$ helicities through their
decays. The study of $\gg \to W^+W^-$  is complimentary to the
corresponding $\epem \rarrow W^+W^-$ channel, but  one can also
check for the presence of anomalous four-point
$\gg \to WW$ interactions  not already constrained
by electromagnetic gauge invariance, such as the effects due to $W^\ast$
exchange.

The  cross section for $\gg \to WW$  at a $TEV$ linear collider
rises asymptotically to a constant because of the spin-one
$t-$ channel exchange:
$\sigma_{\rm asympt}
(\gg \to WW ) \simeq 8\pi\alpha^2/M^2_W \simeq 80$ pb.
This is a rather large
cross section: a linear $\gg$ collider with a luminosity of
10-20 fb$^-1$ will produce of the order of
one million $W^+ W^-$ pairs\refmark\BBW.

A main focus of the pair production studies will be the values of the
$W$ magnetic moment $ \mu_W = {e\over 2m_W}\ (1-\kappa-\lambda)$
and  quadrupole moment $Q_W= - {e\over M^2_W}\ (\kappa-\lambda).$
The Standard Model predicts $\kappa=1$ and $\lambda=0,$ up
to radiative corrections analogous to the Schwinger corrections
to the electron anomalous moment. The
anomalous moments are thus defined as $\mu_A = \mu_W-{e\over M_W}$
and $Q_A = Q_W + {e\over M^2_W}.$

\REF\DHG{S. D. Drell and A. C. Hearn, \sl Phys. Rev. Lett.
\bf 16 \rm (1966) 908; S. B. Gerasimov. \sl Yad. Fiz \bf 2 \rm
(1965) 598 [\sl Sov. J. Nucl. Phys. \bf 2 \rm (1966) 430];
M. Hosoda and K.~Yamamoto, \sl Prog. Theor. Phys. \bf 36 \rm
(1966) 426; see also S. J. Brodsky and J.~R.~Primack, \sl
Ann. Phys. (N.Y.) \bf 52 \rm (1960) 315.}

\REF\SJBHILLER{%
S. J. Brodsky and  J. R. Hiller,
\sl Phys. Rev. \bf D46 \rm (1992) 2141.}

\REF\TUNG{W.- K. Tung, \sl Phys. Rev. \bf 176 \rm
(1968) 2127.}

\REF\SJBDRELL{%
S.  J. Brodsky and S. D. Drell,
\sl Phys. Rev. \bf D22 \rm (1980) 2236.}

The fact that $\mu_A$ and $Q_A$ are close to zero is actually
a general property of any  spin-one system
if its size is small compared to its
Compton scale.  For example, consider the Drell-Hearn-Gerasimov
sum rule\refmark\DHG~ for the $W$ magnetic moment:
$
\mu^2_A = \left(\mu-{e\over M}\right)^2 = {1\over\pi}\int^\infty
_{\nu_{th}} {d\nu\over\nu}\, [\sigma_P(\nu)-\sigma_A(\nu)].
$
Here $\sigma_{P(A)}$ is the total photoabsorption cross section
for photons on a $W$ with (anti-) parallel helicities.
As the radius of the $W$ becomes small, or its threshold energy for
inelastic excitation becomes large,
the DHG integral and hence $\mu^2_A$ vanishes.
Hiller and I have recently
shown\refmark\SJBHILLER~ that this  argument can be generalized to the
spin-one anomalous quadrupole moment as well,
by considering one of the
unsubtracted  dispersion relations for near-forward
$\gamma$ spin-one Compton scattering\refmark\TUNG:
$$  \eqalign{
\mu_A^2 + {2t\over M^2_W}\ &
\left(\mu_A+{M_2\over W}\ Q_A\right)^2 = \cr\crr
& {1\over 4\pi}
\int^\infty_{\nu_{th}} {d\nu^2\over (\nu-t/4)^3}\
Im\, (f_P(s,t)-f_A(s,t))\ . \cr}
  $$
Here $\nu = (s-u)/4$.  One again sees that in the point-like
or high threshold energy limit, both
 $\mu_A \rarrow 0,$ and $Q_A\rarrow 0.$  This  result
applies to any spin-one system, even  to the deuteron or
the $\rho.$ The essential assumption  is the existence of the
unsubtracted dispersion relations; \ie, that the anomalous  moments
are in principle computable quantities.

In the case of the $W$, the finite size correction is expected
to be order $m^2/\Lambda^2$, since the  underlying composite
theory should be chiral to keep the $W$ mass finite as  the composite
scale $\Lambda$  becomes large\refmark\SJBDRELL.
Thus the fact that a spin-one system has nearly the canonical values
for its moments signals
that it has a small internal size; however, it does
not necessarily imply that it is a gauge field.

Yehudai\refmark\Yehudai~ has made extensive studies of the effect
of anomalous moments on  different helicity amplitude contributing to
$\gamma\gamma \rarrow W^+W^-$  cross section.
Figure \figY\
shows the differential cross section  for the process
$\gamma\gamma \rarrow W^+W^-$ in units of $\sigma_{\epem \to
\mu^+ \mu^-}$ as a function of center
mass angle for $\lambda=0, 0.1$ and  $\kappa=1, 0.1.$  The
empirical sensitivity to anomalous couplings from $\gamma\gamma$
reactions
is comparable and complimentary to that of
$\epem \rarrow W^+W^-.$

\vskip1cm\leftline{{\bf 5.  Neutral Gauge Boson Pair Production
in Photon-Photon Collisions}}
\vskip.5cm

\REF\JikiaNG{
 \sl Nucl. Phys.\bf B405 \rm (1993) 24;
\sl Phys. Lett. \bf B298 \rm (1993) 224. See also
G. Jikia, these proceedings;
the {\it Proceedings of the Second International Workshop
on Physics and Experiments with Linear Colliders},
Waikoloa, Hawaii (1993),
Ed. F. A. Harris \etal, World Scientific;
and Protvino preprints IHEP  92-91 (1992), 93-37
(1993).
E. E. Boos and  G.V. Jikia
\sl Phys. Lett. \bf B275 \rm (1992) 164.}

As emphasized by  Jikia\refmark\JikiaNG,
pairs of neutral gauge bosons of the
Standard model can be produced in $\gamma\gamma$ reactions
through one loop amplitudes in the Standard Model at a rate which
should be accessible to a high energy linear collider.
For example, the Standard Model one-loop contributions
for the reaction $\gamma\gamma\rarrow Z^0 Z^0$ is shown in Fig.
\figCC.~   The computation uses
the background nonlinear gauge in order to avoid
four-point couplings between the ghost fields and the $W$ fields.
The ghost fields include  Faddeev-Popov ghost fields
as well as scalar $W$ auxiliary fields.

A familiar example of this type of quantum
mechanical process is the production
of large invariant mass $\gg$ final states
through light-by-light scattering
amplitudes. Leptons, quarks, and $W$  all contribute to the  box graphs.
The fermion and spin-one exchange contributions to the $\gg \to
\gg$ scattering amplitude have the characteristic behavior
$\M\sim s^0 f(t)$ and  $\M\sim i\,s f(t)$ respectively.  The latter is
the  dominant contribution at high energies, so one can use the
optical theorem to relate the forward
imaginary part of the scattering amplitude to the total $\gg \to
W^+ W^-$ cross section\refmark\JikiaNG.  The resulting cross section
$\sigma(\gamma\gamma \rarrow \gamma\gamma)$  is of order 20
fb at $\sqrt s_\gg$ , corresponding to 200 events/year at an NLC
with luminosity 10 fb$^{-1}.$

Figure \figEE\ shows the effective cross section for
$\gamma\gamma\rarrow Z^0 \gamma $  at an NLC assuming the
back-scattered laser spectrum\refmark\JikiaNG. The rate for
transversely polarized $Z$ dominates strongly over longitudinal $Z$,
reflecting the tendency of the  electroweak couplings to conserve
helicity. The cross section for $\gg \to Z_T \gamma$ at $\sqrt
s_\epem = 500$ GeV
is estimated by Jikia to be 32 fb, corresponding to 320 NLC
events/year.   As we discuss below, the channel $\gg \to Z_T Z_T$
provides a serious background to Higgs production in $\gg$ collisions.

\vskip1cm\leftline{{\bf 6. Higgs Physics}}
\vskip.5cm

The origin of the electroweak symmetry breaking is one of the most
exciting problems in particle physics to be solved experimentally
by the present or the next generation of high energy colliders.

Within the Standard Model, the electroweak symmetry breaking
is induced by the Higgs mechanism which predicts the existence of
a new fundamental
scalar particle with a mass below about 1 TeV. The experimental
value of the electroweak mixing angle supports, qualitatively, the
hypothesis that the fundamental particles remain weakly interacting
up to the GUT scale, leading to a Higgs mass below 200 GeV in the
Standard Model. The solution of the hierarchy problem arising in
this situation suggests the supersymmetric extension of the Standard
Model, expanding the scalar sector to a spectrum of at least
five neutral and charged Higgs bosons.

If light Higgs bosons do not exist, the $W$ bosons must interact
strongly with each other at energies of more than 1 TeV. Novel strong
interactions may give rise to the formation of resonances in the
mass range above 1 to 2 TeV.

\REF\LOW{%
F. E. Low, \sl Phys. Rev. \bf 120 \rm (1960) 582.}

\REF\TPC{H. Aihara \etal,
\sl Phys. Rev. Lett. \bf 60 \rm (1988) 2355.}

An important advantage of a photon-photon collider is its potential to
produce and determine the properties of fundamental $C=+$
resonances such as the Higgs boson\refmark{\LOW}.
The present-day analog of $\gg\to$ Higgs production is the
production of narrow
charmonium states.  For example, the TPC$\gg$ collaboration at
PEP\refmark\TPC~
has reported the observation of  6 $\eta_c$ events,  which gives
$\Gamma_{\gg}(\eta_c) = 6.4\ {+5.0\atop-3.4}\, KeV.$
Higher luminosity facilities such as CESR or a  B-factory should allow
extensive measurements of $\gg$ physics at the charm threshold.

\REF\one{%
For a review see J. F.~Gunion, H. E.~Haber, G. L.~Kane and
S.~Dawson, ``The Higgs Hunter's Guide", Addison--Wesley 1990.}

\REF\two{%
D. L.~Borden, D. A.~Bauer and D. O.~Caldwell,
SLAC--PUB--5715 and \sl Phys. Rev. \bf D48 \rm (1993) 4018.}

\REF\three{%
J. F.~Gunion and H. E.~Haber, \sl Phys. Rev. \bf D48 \rm (1993) 5109.}

\REF\GunionCP{%
B. Grzadkowski and J.F. Gunion,
\sl Phys. Lett. \bf B294  \rm (1992) 361.}

\REF\four{%
M.~Kr\"amer, J.~K\"uhn, M. L.~Stong and P. M.~Zerwas,
DESY 93--174 [\sl Z. Phys. C \rm {\it in print}$\,$].}

\REF\six{%
J. F.~Gunion and J. G.~Kelly, Preprint UCD--94--20.}

\REF\seven{%
I. F.~Ginzburg, G. L.~Kotkin, S. L.~Panfil, V. G.~Serbo and
V. I.~Telnov, \sl Nucl. Instr. and Meth. \bf 219 \rm (1984) 5.}

In this section we will concentrate on the light Higgs scenario with
masses below 1 TeV\refmark\one.
The light Higgs bosons can be discovered
and their properties can be studied throughout the entire Higgs
mass range at the proton collider LHC and at $e^+e^-$ linear
colliders in the TeV energy range. Within this environment,
$\gamma \gamma$ collisions can be exploited to solve two problems.
(i) The measurement of the {\it $\gamma \gamma$ widths of Higgs
bosons}\refmark{\two,\three}.
Since the $\gamma \gamma$ coupling to neutral
Higgs particles is mediated by charged particle loops, this
observable provides indirect information on the spectrum of
heavy particles and their couplings to the Higgs field. (ii) It is
easy to measure the {\it
external quantum numbers} ${\cal J^{PC}} =
 0^{++}$ of the scalar Higgs bosons\refmark\GunionCP.
 However, it is
 very difficult to verify the negative  parity of the
 pseudoscalar Higgs boson $A^0$ in the supersymmetric extension
 of the Standard Model. In some parts of the ${\cal SUSY}$
 parameter space, the positive and negative parity of states
 can be measured by using linearly polarized photon beams%
\refmark{\four,\six}\  The formation of ${\cal P} = +$
 particles requires the $\gamma$ polarization vectors
                                     to be parallel
 while ${\cal P} = -$
 particles require the polarization vectors to be perpendicular.
 Back--scattering of linearly polarized laser light provides
 high--energy photon beams with a high degree of linear
 polarization\refmark\seven.
\REF\WaiTang{%
Wai-Keung Tang, to be published.}
\REF\eight{%
For a review see P.M.~Zerwas ({\it ed}.),
``$e^+e^-$ Collisions at 500 GeV: The Physics Potential",
DESY 92--123A+B and 93-123C.}
More generally,
one can use polarized photon-photon scattering to study CP violation
in the fundamental Higgs to two-photon couplings\refmark\GunionCP.
In the case of electron-photon collisions, one can use the transverse
momentum fall-off of the recoil electron  in $e\gamma\rarrow e\pri
H^0$ to measure the fall-off of the
$\gamma \to $ Higgs transition form factor and thus check
the mass scale of the internal massive quark and $W$ loops  coupling
to the Higgs\refmark\WaiTang.

\REF\BBC{%
D. L. Borden, D. A. Bauer, and D. O. Caldwell,
UCSB-HEP 93-01 (1993).}

\REF\Boos{%
E. Boos, {\it Proceedings of the Second International Workshop
on Physics and Experiments with Linear Colliders},
Waikoloa, Hawaii (1993),
Ed. F. A. Harris \etal, World Scientific,
and E. Boos, I. Ginzburg, K. Melnikov,
T. Sack, and
S.~Shichanin, \sl Z. Phys. \bf 56 \rm (1992) 487.}

\vskip1cm\leftline{{\bf 7.
The Higgs Particle of the Standard Model}}
\vskip.5cm

The cross section for the formation of Higgs particles in
unpolarized $\gamma \gamma$ collisions,
$$
\sigma (\gamma \gamma \rightarrow H) = {8 \pi^2\over m_H}
\Gamma (H \rightarrow \gamma \gamma)
{m_H \Gamma_{\tot}/ \pi\over (s - m_H^2)^2 + (m_H \Gamma_{\tot})^2}
$$
is determined by the $\gamma \gamma$ width of the Higgs bosons.
The two $\gamma$'s fusing to scalar Higgs bosons have equal
helicities; if polarized $\gamma$ beams are employed, the
cross section is to be multiplied by a factor two. For small
Higgs masses $\lsim$ 200 GeV, the Breit--Wigner coefficient
is very sharp, yet beyond this range the Higgs boson of
the Standard Model becomes quickly wider\refmark\eight.
The experimentally observed cross section is obtained by
folding the basic formation cross section with the
$\gamma \gamma$ luminosity. Typical counting rates
between 100 and 1,000 events per year can be
achieved in these experiments (see Fig.~\Afigd).
Calculations of the effects of broadening due to the back-scattered laser
energy spectrum. are also given in Refs. [\BBC], [\Cheung],
and
[\Boos].

\REF\nine{%
J.~Ellis, M. K.~Gaillard and D. V.~Nanopoulos, \sl Nucl. Phys.
\bf B106 \rm (1976) 292;  A.~Djouadi, M.~Spira, J. J.~van der Bij and
P. M.~Zerwas, \sl Phys. Lett. \bf B257 \rm (1991) 187;
M.~Spira, Thesis, RWTH Aachen 1992;
K.~Melnikov and O.~Yakovlev, \sl Phys. Lett. \bf B312 \rm (1993) 179.}

\REF\ten{%
H.~Veltman, \sl Z. Phys. \bf C62 \rm (1994) 235.}

The Higgs boson decay to $\gamma \gamma$ is mediated by the loops
of all charged particles\refmark\nine.
The form factors
${\cal F}_i$ in
$$
\Gamma (H \rightarrow \gamma \gamma) =
{G_F \alpha^2 m^3_H\over 128 \sqrt{2} \pi^3}
\left  | \sum N^i_{{\cal C}} e^2_i {\cal F}
_i \right |^2
$$
depend on the mass ratios $\tau = m^2_H / 4 m^2_i$ of the Higgs to
the loop-quark masses. Light particles
decouple. However, if the Higgs couplings grow with the mass of the
particles, particles even much heavier than the Higgs
boson do not decouple and the form factor ${\cal F}_i$ approaches
a constant value in this limit. This applies, for instance, to
the contributions of the charged leptons and quarks in a fourth
family with ${\cal SM}$ charge assignments\refmark\three.
Interfering
with the $t, W$ amplitudes, the $\gamma \gamma$ width of
the Higgs boson depends strongly on the presence of the 4th
family particles. Even crude measurements of this width afford a
``virtual'' glimpse of the area beyond the Standard Model at energy
scales much larger than the scales which are accessible directly.

The Higgs particle of the Standard Model decays below $\sim$ 150
GeV primarily to $b \overline{b}$ quarks, above $\sim$ 150 GeV to
$W$ and $Z$--boson pairs\refmark\eight.
Since the cross section for
$W$--pair production in $\gamma \gamma$ collisions is very
large, $Z$ decays are the appropriate decay channel to search
for Higgs particles in the high mass range while top
decays are difficult to extract from the overwhelming
$\gamma \gamma \rightarrow t \overline{t}$
background\refmark\ten.

\REF\eleven{%
M.~Drees, M.~Kr\"amer, J.~Zunft and P. M.~Zerwas,
\sl Phys. Lett. \bf B306 \rm (1993) 371.}

\REF\fourteen{%
G.~Jikia and A.~Tkabladze, IHEP--Protvino Preprint 1994.}

If the Higgs bosons are detected in the $\gamma \gamma \rightarrow
H \rightarrow b \overline{b}$ channel, the main background
channel is the direct continuum $ \gamma \gamma \rightarrow
b \overline{b}$
production\refmark{\two,\four,\eleven,\fourteen}.
Background
events from one--resolved photon events, $\gamma \gamma
\rightarrow \gamma g \rightarrow b \overline{b}$ (and even
more so from two--resolved photon events) can be
suppressed very efficiently by choosing the maximum
$\gamma \gamma$ energy not much larger than the Higgs mass;
in this case the soft gluon distribution damps the background
rate very strongly\refmark{\four,\six}.
Since the cross section for the
direct production of charm quark pairs is 16 times larger
than for bottom quarks, excellent $\mu$--vertex detectors
must be employed to reduce these background events.

\REF\thirteen{%
D. L.~Borden, V. A.~Khoze, W. J.~Stirling and J.~Ohnemus,
Preprint UCD--94--8.}

While the photons in the signal process $\gamma \gamma
\rightarrow  H$ have equal helicities, the continuum background
production $\gamma \gamma \rightarrow b \overline{b}$ proceeds
mainly through the states of opposite $\gamma$ helicities
\refmark{\two,\three,\thirteen,\fourteen};
this is a consequence of chirality
conservation in massless QCD which requires
${\cal J}_z = \pm 1$ for a pair of back--to--back moving $b \overline
{b}$ pairs at high energies,
$$
\eqalign{{d \sigma (\gamma \gamma \rightarrow b \overline{b})\over
d \cos \vartheta}
& =  {12 \pi \alpha^2 e^4_b\over  s_{\gamma\gamma}}
{\beta\over (1- \beta^2 \cos ^2 \vartheta)^2}\cr\crr
&  \times
\cases{1 - \beta^4 & for  ${\cal J}_z = 0 $\cr
\beta^2 [1 - \cos^2 \vartheta] [2 - \beta^2 (1 - \cos^2 \vartheta)]
&  for  ${\cal J}_z = \pm 2$\cr}
\cr}
$$
where $\beta$ is the c.m. velocity of the $b$ quarks and
                                         $\vartheta$ the
c.m. scattering angle. The suppression becomes
less effective if gluon radiation is taken into account.%
\refmark{\thirteen,\fourteen}\
In particular, the cross section for charm--quark
production increases significantly in the ${\cal J}_z = 0$ channel
through gluon radiation. However, employing sufficiently
powerful $\mu$--vertex detectors, the $c$--quark problem remains
under control also in this case (see Fig.~\Afige).

\REF\sixteen{%
M. S.~Berger, \sl Phys. Rev. \bf D48 \rm (1993) 5121;  D. A.~Dicus
and C.~Kao, \sl Phys. Rev. \bf D49 \rm (1994) 1265.}

For masses of more than 150 GeV, the Higgs particles decay
almost exclusively into $W^+W^-$ and $ZZ$ gauge boson pairs.
Since the background cross section $\gamma \gamma \rightarrow
W^+W^-$ is very large, the $WW$ channel cannot be used to
detect heavy Higgs bosons, and we are left with the $ZZ$ decay
channel. However, the cross section of the background
channel $\gamma \gamma \rightarrow ZZ$ has turned out
to be unexpectedly large\refmark{\JikiaNG,\ten,\sixteen}.
This process is
of higher order in the electroweak couplings, and it is mediated by a
box $W$ loop. (An estimate\refmark\JikiaNG\ of the magnitude of the
background cross section, $\sigma(\gg \to Z_T) \sim
Z_T \sim 250$ fb can be obtained simply by
scaling the $\gg \to \gg$ rate
by the coupling ${g^4_{WWZ}/ e^4} \sim 11$.)
In particular for high energies, the production
of transversely polarized $Z$ bosons dominates the longitudinal
cross section associated with the Higgs channel
by several orders of magnitude. As a result, the detailed analysis
presented in Fig.~\Afigf, leads us to conclude that the Higgs
signal can be detected in the $ZZ$ channel for masses
up to about 350 GeV at a 500 GeV $e^+e^-$ linear collider facility.

\vskip1cm\leftline{{\bf 8.
Higgs Particles in Supersymmetric Extensions
of the Standard Model}}
\vskip.5cm

In the minimal version of the supersymmetric extension of the
Standard Model, a spectrum of five Higgs bosons is predicted:
two neutral scalar particles, one neutral pseudoscalar
particle, and two charged particles. The charged particles
can be produced in pairs in $\gamma \gamma$ collisions; the
neutral particles
are produced singly with the transition amplitude built--up
by the scalar, spin 1/2 and the $W$ boson loops. The
pseudoscalar Higgs boson $A^0$ does not couple to the gauge bosons
at the Born level. The $\gamma \gamma$ width of the
lightest of the neutral Higgs bosons $h^0$, with a mass of order
$m_Z$, is insensitive to the contributions of the ${\cal SUSY}$
particle loops, and the dynamics is determined by the $b, t$ quarks
and the $W$ bosons. The heavy neutral Higgs bosons are affected
by the ${\cal SUSY}$ particle loops only if their masses do not
exceed the threshold considerably\refmark\three.

\REF\seventeen{%
A.~Djouadi, M.~Spira and P. M.~Zerwas, \sl Phys. Lett. \bf B311
 \rm (1993) 255;
K.~Melnikov, M.~Spira and O.~Yakovlev, DESY 94--076 [\sl Z. Phys. C.
{\it in print}].\rm }

Besides the measurement of the $\gamma \gamma$ widths of these
Higgs bosons\refmark{\three,\seventeen},
the production of Higgs bosons in
linearly polarized $\gamma \gamma$ collisions can be used
to discriminate the negative parity $A^0$ state from the
scalar positive parity $h^0, H^0$
states\refmark{\four,\six}.  While
the polarization vectors of the two photons must be
parallel to generate $0^{++}$ particles $[{\cal M}^+ \sim
\vec{\varepsilon}_1 \cdot \vec{\varepsilon}_2]$, they must
be perpendicular for $0^{-+}$ pseudoscalar particles
$[{\cal M}^- \sim  \vec{\varepsilon}_1 \times \vec{\varepsilon}_2
 \cdot \vec{k}_\gamma]$.
                                          Since the
$A^0$ Higgs particle does not couple directly to gauge
bosons, the measurement of the parity in $\gamma \gamma$
collisions will be a unique method in areas of the
parameter space where the decay to $t \overline{t}$ pairs
cannot be exploited.

Since only part of the linear laser polarization is transferred
to the high--energy photon beams, it is useful to define the
polarization asymmetry
$$
{\cal A} = {N^{\parallel} - N^\perp\over
            N^{\parallel} + N^\perp}
$$
where $N^{\parallel}$ and
                $N^\perp$ denote the number of $\gamma \gamma$
events with the initial laser polarization being parallel
and perpendicular, respectively. It follows that ${\cal A}
(0^\pm) = \pm {\cal A}$. For resonance production, the
asymmetry ${\cal A}$ can be expressed by the
appropriate luminosity factors, ${\cal A} = \VEV{\Phi_3 \Phi_3}
/\VEV{\Phi_0 \Phi_0}$, where the third component of the Stokes
vector is defined as $\xi_3 = \Phi_3 / \Phi_0$. The
maximum sensitivity ${\cal A}_{\max} = (\xi^{\max}_3)^2$ is reached
for small values of $x_0 \lsim 1$ and near
the maximum $\gamma \gamma$ energy
(see Fig.~\Afigg)\refmark{\four,\six}.

The background continuum production $\gamma \gamma \rightarrow
b \overline{b}$ dilutes the asymmetries. Because
$N^{\parallel} = N^{\perp}$
in the continuum sufficiently above the threshold, the background
process does not affect the numerator of the asymmetry, yet it does
increase the denominator significantly, thus diminishing the observed
asymmetry. Large integrated $\gamma \gamma$
luminosity of 20 to 100 $fb^{-1}$ are necessary in
general to reach statistically significant conclusions.%
\refmark{\four,\six}\

Under these conditions the polarization asymmetry of the
lightest neutral Higgs boson $h^0$ can be measured
throughout  the relevant parameter range, except
presumably in the very low mass range, Fig.~\Afigh. The
measurement of the parity in the unique case of the
pseudoscalar $A^0$ Higgs particle appears feasible
throughout most of the parameter range below the top
threshold, Fig.~\Afigi. Optimization procedures in
choosing the $e^\pm$ beam energy and the laser frequency
as well as the analysis of other than $b \overline{b}$
decay channels ameliorate this picture
further\refmark{\four,\six}.

\vskip1cm\leftline{{\bf 9. Studying $\bold{WW}$ Collisions
at a Photon Linear Collider}}
\vskip.5cm

One of the most interesting potential applications of photon-photon
collisions at a
high energy linear collider
is  $W W$ scattering,   as illustrated in Fig. \figEEE.

\REF\Zerw{%
Higgs production  in electron-photon collisions is discussed
by K.  Hagiwara
I.~Watanabe, and P. M.  Zerwas,
\sl Phys. Lett. \bf B278 \rm (1992) 187.}

\REF\BB{%
M. Baillargeon and F. Boudjema,
\sl Phys. Lett. \bf B317 \rm (1993) 371.
See also M.~Baillargeon and F. Boudjema,
ENSLAPP-A-400-92, Aug 1992, published
in the {\it Proceedings of the conference
Beyond the Standard Model III}, Ottawa, Canada, (1992) 387.
}

In this process\refmark{\BrodskyW--%
\CheungWWWW}~   each photon is resolved as a $W W$ pair.
The interacting  vector bosons can then
scatter pair-wise or annihilate; \eg,
they can annihilate into a Standard
Model Higgs boson\refmark{\Zerw,\BB}~ or a pair of top quarks.
In principle, one can use this process
as a nearly background-free laboratory for
studying $W W$ interactions.  The scattering reaction leads to two
$W$'s emerging at large transverse momentum in the final state
accompanied by two $W$'s at $p_T \sim M_W$ focussed along
the beam direction.
We can estimate the cross section for $\gamma\gamma
\rarrow WWWW$ to be of order
$
\sigma_{\gamma\gamma\rarrow WWWW} \sim
\left(\alpha\over\pi\right)^2\log^2 {s\over M_w^2} \, \sigma
_{WW\rarrow WW}
$

The splitting function for $\gamma\to W^+ W^-$ is relatively flat
for some $W$ helicities,
so that one has a relatively high probability for
the $W$'s to scatter or annihilate
with a high fraction of the parent $\gg$ energy.
The polarization of the colliding photons
will also provide a critical tool in analyzing the experimental
signals.
One can also hope to utilize the fact that
the polarization of the spectator $W's$ are correlated with
their transverse momentum distributions and longitudinal
momentum fractions.

The colliding $W's$ produced in a photon-photon collider
can interact in many ways, including
photon, Z, W, and Higgs exchange
interactions  in the $t$-  and $s$-channels.
The  identical $W's$ also scatter via u-channel amplitudes.
The  oppositely-charged $W$'s, can  annihilate through a virtual
photon or $Z^0$ or Higgs boson to final states such as
$t\bar t.$  When the $W^+$ and $W^-$ annihilate into a
Higgs boson,
the $\gg \to WWWW$ process is equivalent to $\gg \to W^+W^- H \to
WWWW$, which has been studied, \eg, in Ref. [\BB].
As shown in Fig. \figAA~ the $\gg \to WWH$ production rate  for
a Higgs of mass 100 GeV is of order 0.4 pb at a 2 TeV $\gg$ CM energy.

Even if the Higgs state does not exist, the cross section for $W W$
collisions are still significant at $TeV$ energies.
Because of the equivalence theorem, longitudinally
polarized $W$'s  inevitably become strongly interacting at $TeV$
energies.  The $\gamma\gamma \rarrow WWWW $
and $\gamma\gamma \rarrow WWZZ$ cross sections thus become
maximally large.   An interesting example of electroweak
symmetry breaking occurs when a techni-$\rho$
couples to the  $ZZ, ZW$ and $WW$ channels. Kinematic cuts can be
designed to separate the spectator and active vector bosons
and possibly identify their charge state and polarization.

Cheung\refmark\CheungWWWW~ and Jikia\refmark\JikiaWWWW~
have recently begun systematic studies of
the $\gg \to WWWW$ and $\gg \to WWZZ$ channels based on
various models for electroweak symmetry breaking.
The analyses of the Born amplitudes are exact, without reliance on
the effective $W$ approximation.
Some representative total cross sections\refmark\JikiaWWWW~
for four gauge
boson production in $\gg$ collisions are shown in Fig. \figAA.
The cross sections for some other standard $e^+e^-$ and $\gg$
reactions are also shown.\refmark{\BB,\BaillargeonRev}
Note that the $\gg \to WWZZ$ and $\gg \to WWWW$ cross sections
are larger than the corresponding $e^+e^- \to \bar\nu\nu W W$
and $e^+ e^- \to \nu \bar \nu ZZ$ rates in $e^+ e^-$ collisions at
the same available energy.
Jikia also finds that the Higgs boson with a mass up to 700 GeV
should be easily observed in a 1.5 TeV linear collider,
and that the signal for a heavy (1 TeV) standard model Higgs boson
can be observed at a 2 TeV linear collider.
According to Cheung's estimates\refmark\CheungWWWW, the
physics of this fundamental sector
of the standard model could be explored at $\sqrt s_\gg = 2$ TeV with a
$\gg$ luminosity as low as 10 fb$^{-1}.$
Further studies of backgrounds and analyses of photon
luminosities and kinematic cuts are clearly necessary, but one can be
optimistic that  measurements of   $W W$ collisions
will eventually become
viable at high energy linear colliders. A more complete
discussion may be found in Jikia's  and Cheung's
contributions\refmark{\JikiaWWWW,\CheungWWWW}
to these proceedings.

\REF\bkz{%
This section was prepared in collaboration with M. Kr\"amer.}

\REF\walzer{T. F.\ Walsh and P. M.\ Zerwas, \pl{44}{73}{196}.}

\REF\BKTF{%
S. J. Brodsky, T. Kinoshita, and H. Terazawa,
\sl Phys. Rev. Lett. \bf 27 \rm (1971) 280.}

\vskip1cm\leftline{{\bf 10. Photon Structure Functions
at the Next Linear Collider}\refmark\bkz}
\vskip.5cm

\input TABLES
\def\Im{\mathop{{\cal I}\mskip-4.5mu \lower.1ex \hbox{\it m}}}
\def\Re{\mathop{{\cal R}\mskip-4mu \lower.1ex \hbox{\it e}}}
\def\simgt{\rlap{\lower 3.5 pt
\hbox{$\mathchar \sim$}} \raise 1pt \hbox {$>$}}
\def\simlt{\rlap{\lower 3.5 pt
\hbox{$\mathchar \sim$}} \raise 1pt \hbox {$<$}}

\def\zp#1#2#3{\sl Z. Phys.\ \bf C#1 \rm (19#2) #3}
\def\pl#1#2#3{\sl Phys.\ Lett.\ \bf B#1 \rm (19#2) #3}
\def\np#1#2#3{\sl Nucl.\ Phys.\ \bf B#1 \rm (19#2) #3}
\def\prd#1#2#3{\sl Phys.\ Rev.\ \bf D#1 \rm (19#2) #3}

\def\prep#1#2#3{\sl Phys.\ Rep.\ \bf C#1 \rm (19#2) #3}
\def\niam#1#2#3{\sl Nucl.\ Instr.\ and Meth.\ \bf #1 \rm (19#2) #3}
\def\frac#1#2{{#1\over#2}}

\REF\berger{Ch.\ Berger and W.\ Wagner, \prep{146}{87}{1}.}
\REF\storrow{For a summary see J.\ Storrow, {\it Proceedings,
``Two--Photon Physics
          at LEP and HERA''}, Lund 1994; and M. Fontannaz, {\em ibid}.}
\REF\schuler{G.\ Schuler and T.\ Sj\"ostrand, \np{407}{93}{539} and
                  {\it Proceedings,
                  ``Two--Photon Physics from Daphne to LEP200 and
                   Beyond''}, Paris 1994.}
\REF\DeWitt{%
R. J. DeWitt, L. M. Jones, J. D. Sullivan, D. E. Willen, H. W. Wyld,
Jr., \sl Phys. Rev. \bf D19 \rm (1979) 2046.}
\REF\miller{D.\ Miller, {\it Proceedings,
``ECFA Workshop on LEP200''}, Aachen 1986,
          CERN 87-08, and {\it Proceedings,
          ``Lepton--Photon Symposium''}, Cornell
                 1993.}
\REF\bardeen{W. A.\ Bardeen and A. J.\ Buras,
\prd{20}{79}{166}; D21 (1980)
                  2041 (E);
                  M.\ Fontannaz and E.\ Pilon, \prd{45}{92}{382};
              M.\ Gl\"uck, E.~Reya and A.\ Vogt, \prd{45}{92}{3986}.}
\REF\rossi{G.\ Rossi, \pl{130}{83}{105}.}
\REF\antoniades{I.\ Antoniades and G.\ Grunberg, \np{213}{83}{445}.}
\REF\vogt{For a recent discussion of various schemes for the higher
               corrections, including numerical analyses, see A.\ Vogt,
          {\it Proceedings,
          ``Two--Photon Physics at LEP and HERA''}, Lund 1994.}
\REF\footA{%
In higher orders, for instance in the renormalization
scheme $\overline{MS}$, $ \Lambda^{(N_F)}_{\overline{MS}}$
carries the index of the active flavors. Since the data with
the smallest errors had been collected in the past
for $Q^2$ values of ${\cal O}(10$ GeV$^2$), early experimental
analyses\refmark{\berger}\ correspond to $\Lambda^{(4)}_{\overline{MS}}$
with four flavor degrees of freedom.}
\REF\peterson{C.~Peterson, T.~F.~Walsh and P.~M.~Zerwas,
                   \np{299}{83}{301}.}
\REF\footB{%
For off--shell $\gamma$ targets, $\Lambda^2 \ll P^2 \ll Q^2$, the
gluon radiation is reduced, and the structure function
approaches the Born term again with increasing $P^2$.  See also
T.~Uematsu and T.~F.~Walsh, \pl{101}{81}{263}.}
\REF\sasaki{T. Sasaki et al., \pl{252}{90}{491}.}
\REF\withill{E. Witten, \np{104}{76}{445};
                  C. T.\ Hill and G. G.\ Ross, \np{148}{79}{373};
                  M.\ Gl\"uck and E.\ Reya, \pl{83}{79}{98}.}
\REF\laenen{E.\ Laenen, S.\ Riemersma, J.\ Smith and W.L.\ van Neerven,
                 Fermilab--Pub--93/240--T.}
\REF\comment{We give only crude estimates in this section for effective
                  values of $\alpha_s$ and $\Lambda$; the analysis must
                  necessarily be refined by including higher orders to
                  define $\alpha_s$ properly. See Ref. [\vogt].
                  Alternatively, the $\MS$
                  scheme and scale dependence
                  of the photon structure functions can
                  be eliminated
                  in terms of an effective charge defined from a
                  physical observable. See
S. J. \ Brodsky and H. J.\ Lu, SLAC--PUB--6481 (1994).}
\REF\bjorken{J. D.\ Bjorken, SLAC--PUB--5103.}
\REF\ibes{W.\ Ibes and T. F.\ Walsh, \pl{251}{90}{450};
               S.~M.~Kim and T.~F.~Walsh,
               Minnesota Preprint UMN--TH--1111/92.}
\REF\Cordier{%
See  A. Cordier et al, LEP200
   Workshop, CERN 87-08 (1987).}
\REF\bawa{A.~C.~Bawa and M.~Krawczyk and W.~J.~Stirling
               \zp{50}{91}{293};
               A.~C.~Bawa and M.~Krawczyk, \pl{262}{91}{492}.}
\REF\kuraev{ E. A.\ Kuraev,
L. N. Lipatov and V. S. Fadin, Sov.\ Phys.\ JETP
              45 (1977) 199;
         Ya. Ya.\ Balitzkij and L. N. Lipatov, Sov.\ J.\ Nucl.\ Phys.\
                 28 (1978) 822.}
\REF\forshaw{J. R.\ Forshaw and P. N.\ Harriman, \prd{46}{92}{3778}.}

\REF\ginzburg{I.F.\ Ginzburg, G.L.\ Kotkin, V.G.\ Serbo and V.I. Telnov,
                   \niam{219}{84}{5}.}

The photon structure functions measured in deep--inelastic
electron--photon
scattering\refmark{\walzer,\BKTF} are one of the most
interesting testing
grounds for QCD. On one hand, the process is complicated enough
to reveal rich non--trivial structures within the photon;
on the other hand, it is still sufficiently
simple  to allow for a variety of exciting theoretical predictions.
In contrast to the structure function of the proton, the transverse
structure function of the photon is predicted to rise linearly with
the logarithm of the momentum transfer and to increase with
increasing Bjorken $x$ \refmark\walzer.
This is a consequence of asymptotic
freedom which allows for large transverse momenta in the splitting
of a photon into a quark--antiquark pair. It was first shown by
Witten\refmark\Witten~
that the quark--parton prediction is renormalized
by gluon bremsstrahlung in QCD to order unity and that the absolute scale
of the photon structure function is
fixed by the value of the QCD coupling
constant. Thus
both the shape and size of the photon structure function are
determined by perturbative QCD at asymptotic energies. At
presently available
laboratory energies, the perturbative picture
must be supplemented by conjectures on the residual non--perturbative
component of the structure function.
These  novel qualitative features of the photon structure function were
born out by the pioneering $\gamma\gamma$ experiments at PETRA and PEP
\refmark\berger.

The high luminosities expected for LEP200 and  the $e\gamma$ mode of
prospective linear colliders
will promote this fundamental $e\gamma$ process to
an experimental instrument of high precision.
Two problems can be addressed that are of general interest beyond the
specifics of the $e \gamma$ process itself. (i) Since the size
of the photon structure function is strongly affected by gluon
radiation, the process can be exploited to measure the
QCD coupling constant. At sufficiently large x, the {\it evolution} of
the structure function can be used to extract $\alpha_s (Q^2)$ in a
model independent way.
The evaluation of the {\it absolute size} for
this purpose is affected by the non--perturbative remnants
in the hadronic
components of the structure function.
(ii)  Accurate measurements
of the quark and gluon distributions in the photon will allow predictions
for a large number of phenomena in other fields
such as the
production of large transverse momentum jets,
photons and heavy quark states
in $\gamma\gamma$ and $\gamma$N
collisions \refmark{\storrow,\Resolved}.
Even the total $\gamma \gamma$
and photoproduction cross sections on nucleons may be affected
significantly by the perturbative quark--gluon content of the photon
\refmark\schuler.

\vskip1cm\leftline{{\bf 11. Deep--Inelastic
Electron--Photon Scattering}\refmark\bkz}
\vskip.5cm

If the photon fluctuates (see Appendix I)
into a quark--antiquark pair, three different
dynamical regimes can be distinguished, depending on the
transverse momentum $k_\perp$ of the quarks with respect to the $\gamma$
momentum. The transverse momentum determines the lifetime
$\tau_\ast \sim 1/k_\perp$ of the fluctuation in the $q\bar q$ rest frame
and $\tau \sim E_{\gamma} / k^{2}_{\perp}$ in the laboratory frame.
(See Appendix II.)

(i) For $k_\perp \simlt {\cal O}(\Lambda)$, $\Lambda$ being
the QCD scale parameter, the lifetime is long enough for the quark
to travel to the confinement distance and to  eventually  resonate.
Resonances, $\rho, \omega, \varphi$
can form in this situation, building up
to hadronic photon component
$ |\gamma\!>_{\rm RES} =
e/f_\rho |\rho > + e/f_\omega|\omega> + e/f_\varphi |\varphi> + ... $
If this state is probed deep--inelastically, the quark components
have to be added up coherently, resulting in the
Fock decomposition $ |\gamma >_{\rm  RES} \Rightarrow
\sqrt{2}e/f_\rho [+ \frac23 |u \bar u> - \frac13 |d \bar d>
- \frac13 |s \bar s> + ... ] $

(ii) If $k_\perp \simgt {\cal O}(\Lambda)$, the lifetime
becomes too short for the quarks to form a bound state. Instead, a
shower of quarks and gluons develops. The evolution of this shower
is governed by the DGLAP equations modified
by an inhomogeneous source term for quarks that accounts for
the increased probability of the splitting $\gamma \rightarrow
q \bar q$ if the phase space increases\refmark\DeWitt.

(iii) Finally, if $k_\perp$  is very large, the photons couples
in a point-like fashion to the quarks, corrected to ${\cal O}(\alpha_s)$
etc. by loops and hard bremsstrahlung. This domain is described by
the standard rules of perturbation theory.
\bigskip

The cross section for deep--inelastic scattering $e+ \gamma \rightarrow
e + X$ is parametrized by the transverse $F^\gamma_T$ and
the longitudinal $F^\gamma_L$ structure functions, Fig.~\figA (a),
$$
\frac{d\sigma}{dx\,dy} =
\frac{2 \pi \alpha^2 s_{e \gamma}}{Q^4} \left[1+(1-y)^2\right]
\left [2x F^\gamma_T (x,Q^2) + \varepsilon_y F^\gamma _L (x, Q^2)\right]
\ .
 $$
The transverse structure function can be substituted  by the
more familiar structure function
$F^\gamma_2 = 2 x F^\gamma_T + F^\gamma_L$.
The Bjorken variable $x$ and $y$ can be expressed
in terms of the momentum transfer $Q^2$, the invariant hadronic energy
$W$, and the laboratory energies and electron scattering angle,
$$
x = \frac{Q^2}{2 q \cdot p_\gamma} = \frac{Q^2}{Q^2 + W^2}\quad {\rm
and}\quad
 y = \frac{q \cdot p_\gamma}{k \cdot p_\gamma} = 1 - \frac{E'}{E} \,
\cos^2  \frac{\vartheta}{2}\ .
 $$
Since the degree of the longitudinal (virtual) photon polarization is
given by $ \varepsilon _y = 2 (1 - y) / [1 + (1 - y)^2], F^\gamma_2$
comes with a coefficient $\sim (1 - y)$ for small $y$ and is easy
to measure. By contrast, $F^\gamma_L$ comes with a coefficient $y^2$
which is difficult to measure
since $y$ must be restricted to small values
to reject beam--gas background events\refmark\miller.

The area in $Q^2$ and $x$ that can be explored by LEP200 and a
$e^+e^-$ linear collider at a c.m. energy of 500~GeV [LC500] is
characterized by a parallelogram in the ${\cal P} = \{\log Q^2, \log
(1/x - 1) \}$ plane, Fig.~\figB. Below the one--pion line
``$\pi$''  the photon structure functions vanish identically while
the continuum extends from the ``$\pi \pi$'' threshold into the
upper half of ${\cal P}$. The right boundary $Q^2_{\max}$ is set
by experimental counting rates; $Q^2_{\max}$ is in general
much smaller than $s^{\max}_{e\gamma}$. The parallelogram extends to the
left down to $Q^2_{\min}$, identified in Fig.~\figB\ with the limit
below which the application of perturbative QCD becomes doubtful;
similarly the base--line of the parallelogram which corresponds to
$W^2_{\min}$, while the upper boundary is given by the experimentally
accessible $W^2_{\max}$. Rough estimates of $Q^2_{\max}$ and
$x_{\min}$ in the perturbative regime are displayed in Table~1 for
LEP200\refmark\miller\ and LC500. The LC500 machine has been assumed
to operate in the $e \gamma$ mode so that the $Q^2, x$ range is not only
extended by the higher energy but also by the higher luminosity
compared with LEP.
\bigskip
\centerline{Table 1}
\bigskip
\begintable
 & $Q^2_{\max}$ & $x_{\min}$\cr
LEP200 & $10^3$~GeV$^2$ & $10^{-3}$ \cr
LC500  & $10^5$~GeV$^2$ & $10^{-5}$ \endtable
\smallskip
\caption{Rough estimate of the $q^2,x$ values that will be
reached at LEP200 and in the $e\gamma$
mode of an $e^+e^-$ linear collider
at 500~GeV.}
\bigskip

As a result of asymptotic freedom, the photon structure function
$F_2^\gamma (x, Q^2)$ rises with $\log Q^2$ and also with
Bjorken $x$\refmark\walzer.  The rise of the amplitude
$\gamma^\ast \gamma \rightarrow \bar q q$ in $x$
is damped by gluon radiation at
intermediate to large $x$ while the structure function increases
by quark--antiquark pair creation at small $x$\refmark\Witten.
This leading
logarithmic analysis has been extended to $NLO$ \refmark\bardeen\
so that a complete perturbative analysis is available which can be
confronted with experimental data.
A crude estimate of the accuracy expected at
LEP200 is displayed in Fig.~\figC.

In the leading asymptotic solution spurious singularities
are encountered at $x \rightarrow 0$. They are induced by
poles in the moments of the structure function whenever the
anomalous dimensions $d^j_n (j = \pm, NS)$ approach the
values $l=-1,0,+1...$\refmark{\rossi}\  These poles are due to infrared
singularities at ${\cal O}(\alpha_s^l)$ which are cancelled by
perturbative infrared singularities at the real $\gamma$ vertex.
A scheme for the regularization of these singularities has been
designed in Ref. [\antoniades]; {\it in praxi}
the residual effects due to the regularization at ${\cal O}(\alpha^{-1}
_s)$ and ${\cal O}(1)$ are confined to small $x$ values $\simlt 0.15$.

Since in the present context we are only interested in gross
features of the photon structure functions, we will
illustrate the main points in the framework of the
LO GLAP equations\refmark\vogt\  for the moments of the parton densities
in the photon, $q = \int dx\, x^{m-1} q (x,Q^2)$ etc.,
$$ \eqalign{
\frac{\partial q}{\partial t} &= e^2_qd_B + \frac{\alpha_s(t)}{2\pi}
\left [ A_{qq}*q + A_{qg}*G \right ] \cr\crr
\frac{\partial G}{\partial t} &= \hphantom{e^2_qd_B +}
\ \  \frac{\alpha_s(t)}{2\pi} \left [ A_{gq}*q + A_{gg}*G \right ] \cr}
 $$
where $t = \log Q^2/ \Lambda^2$ and
$\alpha_s (t) = 1 / bt$\refmark\footA.

The structure function is given, as usual, by
$$
F_2^\gamma(x,Q^2) = 2\sum_{fl}e_q^2xq(x,Q^2)
$$
the sum running over the light quark species $u,d,s.$ The quantities
$d_B$ etc. are defined in Ref. [\peterson].
For the non--singlet component $q = q_{2/3} - q_{1/3}$, the
difference of up and down type parton densities, the
solution for the evolution $t_0 \rightarrow t$ can be
cast into two different forms,
$$ \eqalign{
q(t) &= q(t_0)\left[\frac{\alpha_s(t)}{\alpha_s(t_0)}\right]^{d_{NS}}
         + \frac13 \ \frac{d_B}{1+d_{NS}}\left\{t-\left[
         \frac{\alpha_s(t)}{\alpha_s(t_0)}\right]^{d_{NS}}t_0\right\}
\cr\crr
     &= \left\{q(t_0)-q_{pt}(t_0)\right\}\left[
         \frac{\alpha_s(t)}{\alpha_s(t_0)}\right]^{d_{NS}}
   + \frac13\ \frac{d_B}{1+d_{NS}}\frac{2\pi/b}{\alpha_s(t)}\ .\cr}
$$
The last term
$$
q_{pt}(t) = \frac13\ \frac{d_B}{1+d_{NS}}\frac{2\pi/b}{\alpha_s(t)}
 $$
denotes the ``point-like component'', renormalized by the
anomalous dimension $d_{NS}$ to ${\cal O}(1)$ with respect to the
quark--parton term $\frac{1}{3} d_B \log Q^2 / m^2_q$\refmark\footB.
The difference between the full solution and the point-like
component, $ q - q_{pt}$, will suggestively be referred to as the
``hadron--like component'' of the photon.

While the hadron--like component approaches zero asymptotically
for $1 + d_{NS} > 0$, the point-like component grows
$\sim \log Q^2$. This is a consequence of asymptotic freedom
which damps the gluon radiation. Indeed, for a fixed coupling
constant $\alpha_\ast$, the quark density would approach a
finite limit $q \sim \alpha^{-1}_\ast$ asymptotically%
\refmark{\peterson}, Fig.~\figD. The data\refmark\sasaki~
at presently available values of $Q^2$ are
compatible with a linear rise in $\log Q^2$, Fig.~\figE.

The most  characteristic behavior  of the photon
structure function $F_2^\gamma(x,Q^2)$ in QCD
is its continuous linear rise  of with
$\log Q^2$ at fixed $x$.  As emphasized in Ref. [\peterson],  the
fact that this tree graph behavior is preserved to
all orders in perturbation theory is
due to the balance in QCD between the
increase of the phase space for gluon emission
in the scattering processes versus
the decreasing strength of the gluon coupling due to asymptotic
freedom. Although  the logarithmic rise of the
Born approximation result is preserved,  the shape
in $x$ is modified by the QCD radiation.

The heavy $c, b$ and $t$ quark contributions are added as Born
terms plus the standard QCD loop corrections\refmark{\withill,\laenen}.
Since $ x \le x_{\max}
= Q^2/(Q^2 + 4m^2_q)$, the heavy quark threshold traverses the
entire $x$ range from low to very high $Q^2$. By measuring the cross
section for tagged $c$--quark production in on-shell $\gamma \gamma$
collisions, it can be checked experimentally whether
the perturbative QCD calculations\refmark\eleven\
are trustworthy already for
$c$ quarks.

A   comparison
of the QCD predictions given by Laenen \etal\refmark\laenen~ with data
from the PLUTO experiment at PETRA for the photon structure
function  at $\VEV{Q^2} = 5.9$ GeV$^2$ is shown in Fig. \figI.
The underlying contribution due to charm at
leading and higher order is also shown. The shape and
normalization of the structure functions  predicted by
PQCD appears to be consistent with
experiment although the detailed results
depend on the assumed  shape of the photon's  gluon
distribution.
\bigskip

\noindent
{\it 11.1 \ Measurement of $\alpha_s (Q^2)$}\brk
For high moments, or equivalently high values of $x$, the parton
densities are determined by the valence quark distributions in the photon
which for light quarks come in the ratio $u : d : s = 4 : 1 : 1$. In
this range the evolution equations can be applied
directly to $F^\gamma_2$,
$$
Q^2 \frac{\partial F^\gamma_2}{\partial Q^2} =
 \frac{4}{9} d_B -
 \frac{d_{NS} F^\gamma_2 + {\cal O}(G/q)}{\log Q^2 / \Lambda^2}
  $$
with the solution
$$
F^\gamma_2 (Q^2) = F^\gamma_2 (Q^2_0) \left[\frac{\log Q^2_0 / \Lambda^2}
{\log Q^2 / \Lambda^2}\right]^{d_{NS}} + \frac{4}{9}\  \frac{d_B}
{1 + d_{NS}} \left\{\log \frac{Q^2}{\Lambda^2} -
\left [\frac{\log Q^2_0/\Lambda^2}{\log Q^2/\Lambda^2}\right]^{d_{NS}}
\log \frac{Q_0^2}{\Lambda^2}\right\}\ .
 $$
In analogy to the proton structure function the sensitivity of
$F^\gamma_2$ to $\Lambda$ is due to the onset of the
asymptotic behavior \refmark{\comment}.
This is a consequence of the fact that
$F^\gamma_{2 pt} (Q^2) - F^\gamma_{2 pt} (Q^2_0) \sim
 \log Q^2/Q^2_0$ is independent of $\Lambda^2$. The
 sensitivity of the slope $Q^2 \partial F^\gamma_2 / \partial Q^2$ to
 $\Lambda$ is illustrated in Fig.~\figF a. If
$\Lambda^{(4)}_{\overline{MS}}$ can be determined within an error of
$\pm$ 150~MeV, the error on the reference coupling $\alpha_s (m^2_Z)$
will be about 8$\%$.

The  relative strength of the
hadron-like component of photon structure
function approaches zero asymptotically, and the absolute magnitude of
$F^\gamma_2$ is given by the $\Lambda$ parameter in the
point-like component,
$$ \eqalign{
F_2^\gamma(Q^2) &= \left\{F_2^\gamma(Q_0^2)
- F_{2\,pt}^\gamma(Q_0^2)\right\}
\left[\frac{\alpha_s(Q^2)}{\alpha_s(Q_0^2)}\right]^{d_{NS}}
+F_{2\,pt}^\gamma(Q^2)\cr\crr
F_{2\,pt}^\gamma(Q^2) &= \frac{4}{9}\ \frac{d_B}{1+d_{NS}}
\log\frac{Q^2}{\Lambda^2} \cr}
 $$
For any foreseeable energies however the hadron-like component does
affect the absolute size of $F^\gamma_2$ \refmark{\bjorken}. Theoretical
estimates of the hadron-like component span between two extreme
hypotheses. On one side
it has been assumed (see Ref. [\vogt])                     all
that for
sufficiently small $Q^2_0 = {\cal O}(1$~GeV$^2)$ the entire
structure function $F^\gamma_2 (Q^2_0)$ is given by the
VDM contribution. In this case there is little sensitivity to
$\Lambda$, for the same reason as before. In the other extreme
case the perturbative effects are assumed to prevail down to
$Q_0 \simgt {\cal O}(\Lambda)$ and the difference between
$F^\gamma_2 (Q^2_0)$ and $F^\gamma _{2pt} (Q^2_0)$ is
attributed to the VDM component of the photon. If this
difference is indeed correctly described by the
$\rho, \omega, \varphi$ vector mesons, say to within $\pm$ 50$\%$,
the measurement of the photon structure function at $Q^2 \sim
100$~GeV$^2$ can be used to determine $\Lambda$ with an
error of about $\pm$ 120 MeV, Fig.~\figF b. Since the $x$ and
$Q^2$ dependence of $F^\gamma_2 (x, Q^2)$ are predicted,
this hypothesis can be scrutinized experimentally. Taking
the photon target slightly off--shell, another check is
provided by the $P^2$ dependence of the VDM contribution \refmark{\ibes}.
\bigskip

\noindent
{\it 11.2 \ $q/g$ decomposition of the photon}\brk
The {\it quark} decomposition of the photon cannot be
disentangled in inclusive measurements at low energies, since
only the sum of the parton densities weighted by their
electric charges, $F^\gamma_2 = 2 x
[\frac{4}{9} u + \frac{1}{9} d + \frac{1}{9} s + ...]$, is accessible
this way. Apart from the difficult analyses of baryons in
the current jet, the semi-inclusive measurement of mesons,
built--up by $s$ quarks, can be used to isolate the
$q_{1/3}$ component: $K^0 = (d \bar s),
\varphi = (s \bar s), ... $

Increasing $Q^2$ to $\sim 10^3$~GeV$^2$ and beyond, the virtual
$\gamma$ exchange is supplemented by $Z$--boson exchange
\refmark{\Cordier}, accounted for by substituting
$$
e^2_q \rightarrow \frac{1}{4} \sum_{i, j = L, R}
\left [ e_q - \frac{q^2}{m^2_Z + Q^2} \frac{Z_i (e) Z_j (q)}
{\sin_2 \vartheta_W \cos^2 \vartheta_W}\right]^2
 $$
where the electroweak $Z$ charges are given by $
Z_L (f) = I_{3L} (f) - e_f \sin^2 \vartheta_W$ and $
Z_R (f) = - e_f \sin^2 \vartheta_W $
for the left-- and right--handed $Z$ couplings. A quantitative
analysis of the size of this effect is presented in Ref. [\Cordier].

In addition to these ${\cal NC}$
mechanisms,  the charged current processes
$e^\mp + \gamma \rightarrow \nu + x$, Fig.~\figA (b), become important
at high energies. Since the virtual $W^-$ boson can be
absorbed only by $u, \bar d, \bar s ...$ quarks,
and $W^+$ by $\bar u, d, s ...$, the cross sections
are given, both, by
$$
\sigma (e \gamma \rightarrow \nu X)  =
\frac{G^2_F s_{e\gamma}}{2 \pi} \int dx\,dy
\left[\frac{m^2_W}{m^2_W + Q^2}\right]^2 x [u + (1 - y)^2 (d + s) + ...]
 $$

The total cross sections, as well as the $y$ dependence,
provide a combination of parton densities different from the
${\cal NC}$ process so that the $q_{2/3}$ and $q_{1/3}$ densities
can in principle be disentangled. For LEP200 the cross
sections are small, $\sim$ several $10 fb$; they
increase however to ${\cal O}(1 pb)$ at LC500 producing
several thousand charged current events. Since the Weizs\"acker--%
Williams $\gamma$ spectrum is continuous and soft,
the analysis of the $y$ dependence would be very
difficult at LEP200, it is however easy  in
the back-scattered laser $e\gamma$ mode at LC500 where the $\gamma$
spectrum peaks at high energies for polarized beams.

The {\it gluon} component of the photon can be
measured in deep inelastic $e \gamma$ scattering only
indirectly. In analogy to $eN$ scattering, the
evolution of the $\gamma$ structure function
$F^\gamma_2 (Q^2)$ is affected by the gluon density in
the following way,
$$
t \frac{\partial F^\gamma_2}{\partial t}  =
F^\gamma_{2\, box} - d_{NS} F^\gamma _2 +
\frac{(2 \sum_{fl} e^2_q) Aqg}{2 \pi b} G  \ .
 $$

This method appears to work well for proton targets and it
should also provide valuable information on the gluon
content of the photon in particular for low $x$ values.
Other measurements of $G$ are based on jet production
in resolved $\gamma$ processes at HERA \refmark{\storrow},
inelastic Compton scattering mediated by resolved photons
\refmark\bawa, as well as photoproduction of heavy quark states
\refmark{\Resolved}.
\bigskip

\noindent
{\it 11.3 \ Small--x phenomena}\brk
The fact that in the basic photon splitting process $\gamma
\rightarrow q \bar q$ the transverse momentum is not
limited leads to two interesting questions: (i) The
strong ordering along the gluon chain in the standard DGLAP
analysis associates high energy jets in the evolution with
small transverse momenta.
In contrast, this rule does not apply to
the perturbative pomeron in the BFKL domain at very
small $x$ for moderate $Q^2$ \refmark\kuraev.   It is therefore an
interesting experimental problem to investigate the
correlation between jet energies and transverse momenta
in deep inelastic scattering $e\gamma\to e+ {\rm  jets}$,
exploiting the difference between the initial parton configurations
in the photon and the nucleon.
(ii) Large transverse momenta in $\gamma \rightarrow
q \overline{q}$ correspond to a small transverse size
of the $(q \bar q)$ pair in space. This leads to
a high density of the gluons emitted subsequently which is
eventually stopped by screening effects. Screening
effects associated with the standard confinement radius $R$
of the hadronic $\gamma$ component have been analyzed in
Ref. [\forshaw].

\vskip1cm\leftline{{\bf 12. Polarization
Effects in $\bold{\gg}$ Collisions}\refmark\bkz}
\vskip.5cm
Weizs\"acker--Williams photons, as well as back--scattered
laser photons, can be generated in a variety of polarization
states. Weizs\"acker--Williams photons are automatically
polarized linearly in the production plane. If the initial
$e^\pm$ beams are circularly polarized, part of the
polarization is transferred to the photons.
(See Section 2.) The degree
of polarization is determined in both cases by the fraction
$\zeta$ of energy transmitted from the $e^\pm$ beams to
the photons,
$$ \eqalign{
{\rm (i)\ linear\ polarization} &: \varepsilon(\zeta)_{WW} =
                                  2(1-\zeta)/[1+(1-\zeta)^2] \cr
{\rm (ii)\ left--right\ asymmetry} &: A(\zeta)_{WW} =
                    (2-\zeta)\zeta/[1+(1-\zeta)^2]\ .\cr}
 $$
For back--scattering of laser photons\refmark{\ginzburg},
the spectrum is given by
$dN/d\zeta \sim \phi_0 + \lambda_e\lambda_\gamma\phi_1$, with
$$ \eqalign{
\phi_0&= \frac{1}{1-\zeta} + 1 - \zeta - 4r(1-r) \cr\crr
\phi_1&=  x_0r(1-2r)(2-\zeta) \cr}
 $$
and $r=x_0^{-1}\zeta/(1-\zeta) \le 1$;
$x_0$ being the square of the invariant
$(\gamma e)$ energy in units of the electron
mass which in general is chosen
between $\sim 1$ and $\sim 4$. The degree of
polarization of the high energy
$\gamma$ beam follows from
$$ \eqalign{
{\rm (i)\ linear\ polarization}
&: \xi_3 = \phi_3/\phi_0 \quad {\rm with} \quad \phi_3=2r^2\cr\crr
{\rm (ii)\ left--right\ asymmetry}
&: A=(\lambda_e\phi_4+\lambda_\gamma\phi_5)/
             (\phi_0+\lambda_e\lambda_\gamma\phi_1)\cr\crr
     &\vphantom{A=} {\rm with}\
        \phi_4 = x_0r[1+(1-\zeta)(1-2r)^2]   \cr\crr
     &\vphantom{A=} {\rm  with}\
        \phi_5 = (1-2r)[1/(1-\zeta)+1-\zeta]  \ . \cr}
 $$
Spectra and asymmetries are displayed for the two cases in
Fig.~\figH.
\bigskip

\REF\pewazer{C.\ Peterson, T. F.\ Walsh and P. M.\ Zerwas,
                  \np{174}{80}{424}.}
\REF\altarelli{G.\ Altarelli and G.\ Martinelli, \pl{76}{78}{89}.}

\noindent
{\it 12.1 \ The Longitudinal photon structure function}\brk
Massless quarks which absorb a longitudinally
polarized virtual photon must
have non--zero transverse momenta.
This is a consequence of angular momentum
conservation since the $\gamma q\bar q$ vertex is helicity--conserving.
Non--zero transverse momenta are generated in the point-like
$\gamma\to q\bar q$ splitting process\refmark{\pewazer}\
and by gluon radiation
off the quark beam\refmark\altarelli.
For the first mechanism the absorption
rate is of order $\int dk_\perp^2/k_\perp^2 \cdot k_\perp^2/Q^2
= {\cal O}(1)$,
in the second case $\sim N(Q^2)\alpha_s(Q^2)=
{\cal O}(1)$. Both contributions
are therefore scale--invariant in leading order.
The quark--parton diagram
provides the dominant contribution, for light quarks
$$ \eqalign{
F_L(x,Q^2) &= \frac{12\alpha}{\pi}\sum_{fl}e_q^4x^2(1-x)\cr\crr
  &\quad  + \frac{\alpha_s(Q^2)}{2\pi}\int dy\,dz\  \delta_1(x-yz)z^2
    \left[\frac83 F_2^\gamma + \frac43<e^2_q>yG(1-z)\right]
\  ,    \cr}
 $$
to which the contribution of heavy quarks,
corrected by gluon loops and gluon radiation, must be added%
\refmark{\withill, \laenen}.

A linear collider can also provide a clear and simple test
of QCD in the case where both electrons are tagged at large
momentum transfer so that both photons are virtual. The leading
contributions to the photon structure functions take on the point-like
form characteristic of  the direct photon couplings to the
quarks.
\bigskip

\noindent
{\it 12.2 \ Linear $\gamma$ polarization}\brk
The linear $\gamma$ polarization
gives rise to an azimuthal asymmetry of the
$[e,e']$ scattering plane with respect to the polarization vector,
$d\sigma/d\phi \sim \cos2\phi \cdot
\varepsilon F_{{\cal X}}(x,Q^2)$. In the
quark--parton model \refmark{\pewazer}\
the structure function $F_{{\cal X}}$
is scale invariant,
$$
F_{{\cal X}} = -\frac{3\alpha}{\pi}\sum e_q^4 \cdot x^3\ ,
$$
modified to ${\cal O}(\alpha_s)$ by QCD corrections.
\bigskip

\REF\efremov{A. V.\ Efremov and O. V.\ Teryaev, \pl{240}{90}{200};
                  S. D.\ Bass, Int.\ J.\ Mod.\ Phys.\ A7 (1992) 6039;
                  S.\ Narison, G. M. Shore and G.\ Veneziano,
                  \np{391}{93}{69};
            G. M. Shore and G.\ Veneziano, Mod.\ Phys.\ Lett.\ A8 (1993)
                  373.}

\noindent
{\it 12.3 The  polarized photon structure functions}\brk
As in the case of $\vec e\vec p$ scattering, the spin structure function
$g_1^\gamma(x,Q^2)$ of the photon is measured by the asymmetry of
right/left--polarized electrons scattered off
polarized targets. This structure
function has been recognized as a
very interesting physical observable
\refmark\efremov, which is
deeply connected with the chiral properties of QCD.
For sufficiently large $Q^2$, the spin structure function
$g_1^\gamma(x,Q^2;P^2)$, $P^2$ denoting the mass squared of
the target photon,
fulfills the following sum rules
$$ \eqalign{
\int_0^1 dxg_1^\gamma(x,Q^2;P^2) = \cases{
 0 & for $P^2=0$\cr
 N_{{\cal C}} \frac{\alpha}{\pi}\sum_{fl}e^4_q\left[1+{\cal O}(
 \log^{-1}Q^2, \log^{-1} P^2)\right] &
 for  $P^2$ large \cr} \ .\cr}
 $$
The first moment of $g_1^\gamma$ is given by the matrix element
$\VEV{\gamma|j_{5\mu}|\gamma}$
which counts the difference between right-- and
left--handed polarized quarks in the photon. This matrix element is
built--up
by two elements, the axial anomaly accounted for by the perturbative
quark--loop contributions, and hadronic
contributions by the Goldstone bosons
associated with the spontaneous breaking
of chiral symmetry. [This discussion
also applies to the singlet axial current
modulo logarithmic corrections.]
Electromagnetic current conservation leads to the cancellation of the two
contributions for $P^2=0$. For $P^2$ large,
on the other hand, the hadronic
contribution vanishes to ${\cal O}(1/P^2)$,
and the non--zero value of the
first moment of the spin structure function
is entirely due to the anomaly.

\vskip1cm\leftline{{\bf 13. Jet  Physics at a Photon Linear Collider}}
\vskip .5cm

The distinction between the direct, versus resolved, hadron-like
contributions to photon interactions becomes especially clear in  two-%
photon jet physics.   If both photons couple directly
to a pair of quarks, the final state is similar to that of $\epem$
annihilation: two co-planar jets are
produced without any source of hadronic
spectators emitted along the beam direction.
Such events would be extraordinarily rare at an the analogous  meson-%
meson collider.

\REF\CHLS{%
C. H. Llewellyn Smith, \sl Phys. Lett. \bf B79 \rm (1979)  83.}

In the case of once-resolved processes, one photon scatters directly
on a constituent quark of the other photon, leaving spectators just in
one beam direction.  In the case of twice-resolved two-photon
processes, jets are produced by  any of the various $q q$ $q g$ and
$g g$ QCD 2 to 2  scattering subprocesses.   Despite their markedly
different origins,  the cross sections for these two photon jet
production processes are  all scale invariant in leading order
\refmark{\Resolved,\CHLS}; that is, in leading logarithm
approximation, they each have the form:
$$ {d\sigma\over d^3p/E}\ (\gamma\gamma\rarrow {\rm Jet} +X) =
{\alpha^2\over p^4_T}\ F(x_T,\theta_{cm}) \ .
 $$
The logarithmic fall-off of the subprocess cross section
is precisely compensated by the  increasing  strength of the resolved
photon structure function.
The $x_T= 2 p^{\rm jet}/{\sqrt x} $ dependence of
$F(x_T,\theta_{cm})$  has a power-law fall-off at large $x_T$:  $\sim
(1-x_T)^N$  where  the index $N$  can be
computed at $x \sim 1 $ simply by counting the number of beam
spectators\refmark\Resolved.

\REF\Tauchi{T. Tauchi,\etal,
{\it Proceedings of the Second International Workshop
on Physics and Experiments with Linear Colliders},
Waikoloa, Hawaii, (1993),
Ed. F.~A.~Harris \etal, World Scientific.
H. Hayashii, \etal, KEK-Preprint-93-47, (1993).}

\REF\LAC{H. Abromowicz, K. Charchula, and A. Levy,
\sl Phys. Lett. \bf B269 \rm (1991) 458.
A. Levy, {\it Proceedings of the IXth International
Workshop on Photon-Photon Collisions},
D. Caldwell and H. Paar, eds. (World Scientific, 1992.)}

An illustration of the various contributions to the jet transverse
momentum distribution from direct, single and  twice resolved
contributions as calculated by Drees and Godpole
is shown in Fig. \figL\refmark\Resolved.  The dotted curve
shows the background from $e^+ e^-$ annihilation events with single
hard photon radiation from the initial state.
A recent comparison of these  predictions for single jet and
two jet processes with TRISTAN data\refmark\Tauchi~
obtained from thrust and
other jet variable analyses appear to confirm the
presence of both direct and
resolved contributions, although
there are uncertainties from higher order corrections to the jet rate
normalization and the assumed form of the photon's gluon
distribution. The largest uncertainty in these results
is due to the unknown form
of the gluon distribution within
the resolved photon\refmark{\Resolved,\LAC}.

\vskip1cm\leftline{{\bf 14. Contribution  from Mini-Jets to the
$\bf{\gg}$  Cross Section\brk}}
\vskip .5cm

\REF\FS{%
J. K. Storrow, {\it Proceedings of the Second International Workshop
on Physics and Experiments with Linear Colliders},
Waikoloa, Hawaii, (1993),
Ed. F.~A.~Harris \etal, World Scientific.
J. R. Forshaw and J. K. Storrow,
{\it Proceedings of the IXth International
Workshop on Photon-Photon Collisions},
D. Caldwell and H. Paar, eds. (World Scientific, 1992.);
\sl Phys. Rev. \bf D46 \rm (1992) 4955.}

\REF\CBP{%
P. Chen, {\it Proceedings of the Second International Workshop
on Physics and Experiments with Linear Colliders},
Waikoloa, Hawaii, (1993);
Ed. F. A. Harris \etal, World Scientific.
P. Chen, T. L. Barklow, M. E. Peskin,
\sl Phys. Rev. \bf D49 \rm (1994) 3209.}

One of the uncertainties concerning  QCD predictions for photon-
photon collisions is the size of the total inelastic cross section  which
can be attributed to from mini-jets, \ie,
jets of $p_T$  beyond a cutoff of order of a few GeV.
Early work by Drees and Godpole\refmark\Resolved~
had suggested that the
production rate for mini-jets  could rise so fast with
energy that mini-jets
would provide a  significant  and troublesome minimum
bias background to
the study  of  $\epem$ events at an  NLC.  However,   recent analyses
by Forshaw and Storrow\refmark\FS~ and by Chen, Barklow,
and Peskin\refmark\CBP\
have now shown  that the rise of the mini-jet rate is moderate into
the TeV linear collider regime, and that
the resulting backgrounds  to physics
signals  are in fact minimal.

The new analyses\refmark{\FS,\CBP}~
are based on a two-component form for
total inelastic cross sections:  an energy-independent term  $\sigma_0$,
plus a rising PQCD contribution obtained by  integrating $2 \rarrow
2$ QCD processes from $p_{t \min}= 3.2$ GeV to the kinematic limit.
This parameterization is
consistent
with the measured rate of mini-jets measured by UA1,
the energy dependence of
the $p\bar p$ cross section, as well as $\sigma_{\gamma p}(s)  $
determined by  the ZEUS collaboration at HERA. The cross section for
mini-jets must be unitarized so that
the integral of the cross section $d\sigma/dy$ is  normalized to
inelastic cross section times the average multiplicity of mini-jets
$\VEV{n}.$  As shown in  Fig. \figO\refmark\FS,
the eikonalization of the subprocess cross section leads to a significant
reduction in the predicted value and a rise with energy of the
$\gamma\gamma$ inelastic cross section.
The net result for the number of jets with $p_T >
5\ GeV$ produced per crossing at an NLC is only of order 5  to  8
$\times 10^{-2}$
for typical linear collider designs.

\REF\GIS{%
I. Ginzburg, {\it Proceedings of the Second International Workshop
on Physics and Experiments with Linear Colliders},
Waikoloa, Hawaii, (1993),
Ed. F. A. Harris \etal, World Scientific.}

The physics of unitarization has been analyzed from a different
perspective by Ginzburg, Ivanov and Serbo\refmark\GIS.   In
the regime $s_\gg >>p^2_T > \mu^2,$ where  $\mu$ is the
confinement scale of QCD, one can apply perturbative QCD to compute
the set of gluon exchange chains representing the QCD Pomeron,
as in Lipatov's well-known work.
It is also possible to compute the double diffractive contribution
where two gluon chains to
recombine leaving a rapidity gap between the
di-jets.  Ginzburg \etal\ argue that eikonalization must be taken into
account
when the rates for these two processes become equal.
With this assumption, they
find that eikonal corrections must be applied to the
perturbative QCD factorized predictions
for the jet production cross section at $\sqrt{s_\gg}=$ 500 GeV  even
at jet transverse momentum as large as $p_{T} = 26$  GeV.

\vskip1cm\leftline{{\bf 15. Single and Double Diffraction
in  Photon-Photon Collisions}}
\vskip.5cm

\REF\GinzburgIvanov{%
I. F. Ginzburg and D. Yu. Ivanov,
\sl Nucl. Phys. \bf B388 \rm (1992) 376.}

\REF\Chernyak{%
V. L. Chernyak and I. R. Zhitnitsky,
\sl Phys. Rep. \bf 112 \rm (1984) 1783.
Novosibirsk Inst. Nucl. Phys. Acad. Sci. preprint
82-44  (1982).}

The high energies of a $\gg$ collider
will make the  study of double diffractive
$\gg \to V^0 V^0$ and semi-inclusive
single diffractive processes $\gg \to V^0
X$  in the Regge regime $s >> |t|$  interesting.  (See Fig. \figWW.)
Here $V^0 = \rho, \omega\phi,J/\psi,\cdots$  If   $|t|$
is taken larger than the QCD confinement scale, then
one has the potential for a
detailed study of fundamental Pomeron processes and its gluonic
composition\refmark{\Ginzburg,\GinzburgIvanov}.
As in the case of large angle exclusive $\gg$ processes, the scattering
amplitude is computed by convoluting the hard scattering  PQCD
amplitude for $\gg \to q \bar q q \bar q$  with the vector meson
distribution amplitudes. As shown by
Chernyak and Zhitnitsky\refmark\Chernyak,  the
two gluon exchange contribution dominates in the Regge regime,
giving  a characteristic exclusive process scaling law of order
$
{d\sigma\over dt}\ (\gamma\gamma\rarrow V^0V^0)\sim
{\alpha^4_s(t) / t^6}.$
Recently, Ginzburg \etal\refmark\Ginzburg\ have shown that  the
corresponding $\gamma\gamma\rarrow $  pseudoscalar
and tensor meson channels can be used to
isolate the Odderon exchange contribution, that is contributions
related at a fundamental level to three gluon exchange.

\vskip1cm\leftline{{\bf 16. Heavy Quark Pair Production in
$\bold{\gg}$ Collisions}}
\vskip.5cm

The leading contributions to heavy quark
production in $\gg$ collisions are
illustrated in  Fig. \figR.  The resolved
contributions depend in detail on the
assumed form of the photon's gluon distribution.

\REF\EboliHQ{
O. J. P. Eboli, M. C. Gonzalez-Garcia, F. Halzen, and S. F. Novaes,
\sl Phys. Rev. \bf D47 \rm (1993) 1889 and
\sl Phys. Lett. \bf B301 \rm (1993) 115.}
\REF\Eboli{%
O. J. P. Eboli, M. C. Gonzalez-Garcia, F. Halzen, and D. Zeppenfeld,
\sl Phys.Rev. \bf D48 \rm (1993) 224. See also  F. Ginzburg and V. G.
Serbo,  {\it Proceedings of the Second International Workshop
on Physics and Experiments with Linear Colliders},
Waikoloa, Hawaii (1993),
Ed. F. A. Harris \etal, World Scientific.}
\REF\DKZZ{%
M. Drees, M. Kr\"amer, J. Zunft, and P. M. Zerwas, DESY 92-169 (1992).}

The cross section for direct heavy quark production
$\sigma \sim {\pi\alpha^2/ m^2_Q}$ is of order 130 nb
and 1300 pb for $ c \bar c$ and $b \bar b$ production, respectively.
The top pair cross section $\sigma_{t\bar t}$
is of order of  $\half \sigma_{\epem
\to \mu^+ \mu^-}$ at the corresponding
energy\refmark{\Resolved,\EboliHQ,\Eboli}.~
Figure \figS \ shows the prediction of Drees, Kr\"amer, Zunft and
Zerwas\refmark\DKZZ\  for
the inclusive charm production cross
sections in $\epem \to \epem q \bar q X$
using the equivalent photon approximation. The vertical bars on the left
represent the estimated uncertainty due
to the scale dependence of the lowest
order of predictions. The bar on the right
shows the dependence on the quark mass.
TPC$\gg$ and TRISTAN data for
$\sigma(\epem\rarrow  \to D^{\ast\pm} X) $
are compared with the QCD prediction of Drees \etal\
in Fig. \figT.

\REF\KMS{%
J. H. K\"uhn, E. Mirkes, and J. Steegborn,
Karlsruhe preprint TTP92-28 (1992).
Z. Phys. C57 (1993) 615.}
\REF\KuhnZerwas{%
J. H. K\"uhn  and P. M. Zerwas,
\sl Phys. Rept. \bf 167 \rm (1988) 321.}
\REF\BKZ{%
S. J. Brodsky, G. K\"opp, and P. M. Zerwas,
\sl Phys. Rev. Lett. \bf 58 \rm (1987)  443.}
\REF\Bigi{%
I. L. Bigi, F. Gabbiani, and V. A. Khoze, SLAC-PUB 5951 (1992). }

Figure \figU\ a
shows the cross section predicted by K\"uhn, Mirkes
and Steegborn\refmark\KMS\  for $ t \bar t$
production including higher order QCD corrections.
The predicted rate is given for $m_t=150$ GeV as a function of the
center mass energy
of the $\epem$ collider, where a convolution with the computed
photon energy spectrum of the back-scattered laser beam is assumed.
Since the
top mass is greater than  120 GeV,  its decay  width due
to weak decays $\Gamma = 2\Gamma(t\rarrow bW)  $ is
so large that true bound states $t \bar t$ cannot form; nevertheless
there can be significant threshold
effects\refmark{\KuhnZerwas,\BKZ}.\
As shown in Fig. \figU b, if
the experiment  resolution in $\M_{t\bar t}$  is sufficient, then it
may be possible to resolve the predicted
structure of the $\gg \to t \bar t$ cross section near the top
threshold. Detailed predictions for this threshold dependence as a
function of the top quark mass have been given by Bigi,
Gabbiani, and Khoze\refmark\Bigi.~
The combination of  $\gg$ $(C=+) $and  $\epem$  $(C=-) $ measurements
of the $ t \bar t$ threshold spectrum could provide a very precise
value for the top quark mass.

\vskip1cm\leftline{{\bf 17. Single Top Quark Production}}
\vskip.5cm

\REF\JikiaTop{%
G. V. Jikia,
\sl Nucl. Phys. \bf B374 \rm (1992) 83.}

A single top quark can be produced in electron-photon collisions at
an NLC through the process $e^-\gamma
\rarrow W^- t\nu$\refmark{\JikiaTop,\Cheung}.
This process can be identified through the $t \to W^+ b$ decay with
$W\ \to \ell \bar \nu.$  The rate  is thus sensitive to the
$V_{tb}$ matrix element  and  possible fourth generation
quarks and anomalous couplings.
An interesting background is the virtual $W$ process
$e\gamma\rarrow W^\ast -\nu \to W^- H \nu,$ where the Higgs
boson decays to $b\bar b$ and $W^-\to \ell \bar \nu.$
The rates for the signal and background processes as a function of
$\sqrt s_\epem$  computed by
Cheung\refmark\Cheung~ and Yehudai\refmark\Yehudai~
are shown in Fig. \figQQ.

\vskip1cm\leftline{{\bf 18. Higgs Plus Top Quark Pair Production}}
\vskip.5cm

The tree process $\gamma\gamma\rarrow t {\bar t} + H$
can provide a direct measure of  coupling of the Higgs boson to heavy
quarks.
The cross section has been estimated by
Boos\refmark\Boos~ and Cheung\refmark\Cheung~ to be of order
1 to 5 fb.
Cheung\refmark\Cheung~ has also computed the radiative corrections to
$\gamma\gamma t\bar t $ from final state Higgs exchange
interactions. The correction is of $\O (2 - 4\%)$ for  typical
values of the  Higgs boson mass and top quark mass.

\vskip1cm\leftline{{\bf  19. Conclusions}}
\vskip.5cm

\REF\super{%
F. Cuypers, G. van Oldenborgh, R. R\"uckl, CERN-TH 6742/92 (1992).
\sl Nucl. Phys. \bf B383 \rm (1992) 45.}
\REF\ee{%
See, for example, P. H. Frampton,
\sl Mod. Phys. Lett. \bf A7 \rm (1992) 2017, and
the contributions of C. Heusch and P. H. Frampton to  the
{\it Proceedings of the Second International Workshop
on Physics and Experiments with Linear Colliders},
Waikoloa, Hawaii (1993),
Ed. F. A. Harris \etal, World Scientific.}

Photon-photon collisions at a linear $\epem$ collider in the TeV
range will provide an extraordinary window
for testing electroweak and QCD
phenomena as well as proposed extensions of the Standard Model, such as
supersymmetry\refmark\super, technicolor,
and other composite models.  Two-photon
physics is unique in that virtually
all charged particles and their bound
states and resonances  with positive $C$ can be produced;
in addition one can
access pairs of fundamental neutral particles through one-loop
corrections. It will be essential to have the capability
of back-scattered laser beams at the
NLC,  since it is expected  that the resulting luminosity and effective
energy of photon-photon and photon-electron
collisions will be comparable to that of the primary $\epem$ collisions.
Such a facility, together with polarized
electron beams, will also allow the study of the physics of
highly-polarized photon-photon and electron-photon collisions.  There is
also a wide range of physics topics which could be addressed in polarized
electron-electron collisions, if such a capability were available%
\refmark\ee.

Two-photon physics is an extensive phenomenological field, having
elements in common with both $\epem$ and hadron-hadron collisions.
However, the combination of direct photon and resolved processes gives
$\gg$ physics an extra dimension in probing new phenomena.  For
example, since each photon can be
resolved into a $W^+ W^-$ pair, high energy photon-photon collisions can
provide a remarkably background-free laboratory for studying
$W W$ collisions and
annihilation. Thus a photon-photon collider can also
become the equivalent a $W W$ collider to study whether the
interactions of longitudinally polarized $W$'s are controlled by Higgs
annihilation and exchange or a new type of strong interaction.

It is clear that $\gg$ collisions are an
integral part of the NLC physics program.
It was possible to highlight only a small part of the possible new $\gg$
physics topics here. At present energies, studies of $\gg$
collisions at CESR, PETRA, PEP, TRISTAN, and LEP  have led to a number of
important tests of perturbative and
non-perturbative QCD in  exclusive and
inclusive reactions, heavy quark phenomena, and  resonance
formation. Reviews of the results of these experiments and the
underlying theory of $\gg$ collisions
are given in Refs. [\Kessler], [\SJB],  [\KolanoskiZerwas],
[\Field], [\Kolanoski], [\SJBshoresh] and
the proceedings of the International
Workshops on Photon-Photon collisions.

\bigskip
{\noindent \bf Acknowledgements}
\bigskip

\noindent
This work was supported in part
by Deutsche For\-schungs\-ge\-mein\-schaft DFG.
Part of this work was also presented at the Second International
Workshop on Physics and Experiments with Linear Colliders,
Waikoloa, Hawaii.
The review of the photon structure function  given here
is based on work done in
collaboration with Michael Kr\"amer. SJB also thanks
David Borden, Michael Boulware, George Jikia, and Ilya
Ginzburg for helpful conversations.
\bigskip

\vskip1cm\leftline{{\bf Appendix I: The Photon's Light-Cone
Fock Expansion}}
\vskip.5cm

One can  distinguish the various contributions to the
photon's direct and resolved interactions in the following way:
Consider
the Lagrangian for the Standard Model cutoff at an ultraviolet scale
$\Lambda$  and the corresponding light-cone time $\tau =
t-z/c$ evolution operator; \ie, the light-cone   Hamiltonian
$H^{(\Lambda)}_{LC}.$
The photon is the physical zero-mass eigenstate of the full Hamiltonian.
Any eigenstate of the full Hamiltonian can be expanded as a sum of
eigenstates  of the free Hamiltonian:
$
\ket\psi = \sum_n \ket n\VEV{n\,|\,\psi} \ .
$
The photon state is thus equivalent to a coherent
sum of free Fock states with the
same charge and color singlet quantum numbers. The coefficients in
this expansion,
$
\VEV{n\,|\,\psi} = \psi^{(\Lambda)}_{n/\gamma}
(x_i,\vec k_{\perp i},\lambda_i),
$
with
$
\sum x_i = 1 \, $ and $ \sum \vec k_{\perp i} = 0
$
are the basic wavefunction matrix elements  needed to describe the
photon in terms of its  quark and gluon and other Standard Model
degrees of freedom. The $\psi^{(\Lambda)}_{n/\gamma}
(x_i,\vec k_{\perp i},\lambda_i),$
are frame independent functions of
the light-cone fractions $x_i = k^+_i/p^+,$
the relative transverse momenta $k_{\perp i},$ and the
spin projections $\lambda_i$\refmark\LepageBrodsky.
Since the photon is an elementary field,
the physical photon has a non-zero bare component in the Fock expansion:
$
\psi^0_{\gamma}(x,k_\perp) =16 \pi^3 \sqrt{Z_3(\Lambda^2)}\delta(1-x)\,
\delta^2(\vec k_\perp)\,  \delta_{\lambda\lambda\pri}
$
where $Z_3(\Lambda^2)$ is the probability
that the photon stays a bare photon at the cutoff scale $\Lambda.$

Given the photon's Fock expansion we can calculate the photon-
photon scattering
amplitude at high momentum transfer in
terms of its constituents' interactions
in the factorized form  shown in Fig. \figDD\
and
$$
\M = \sum_{n,m} \int {\bar\pi d^2k_\perp dx\over
16\pi^3}\
\psi^{(\Lambda)}_n (x_i,k_{\perp i}\lambda_i)
\int {\bar\pi d^2\ell_\perp dy\over 16\pi^3}\
\psi^{(\Lambda)}_m(y_i,\ell_{\perp i},\lambda_i)
T^{(\Lambda)}_{ab\rarrow cd} \ .
$$
Here $T_{ab\rarrow cd}^{(\Lambda)}$ is a sum over all
$2\rarrow 2$ and higher subprocess amplitudes.
It is irreducible and contains all the interactions, radiative
corrections, and
loop corrections with $k_\perp^2$ greater than the separation scale
$\Lambda^2$.
Higher particle processes are generally higher twist
and  thus power-law suppressed at large momentum transfer.  In the
expansion
of the $\gamma \gamma$ scattering amplitude one thus obtains the
direct pair production processes from the bare photon components
as well as the resolved contributions.
One can then square the matrix element, integrate over undetected
variables, and derive the usual factorized form for hard
scattering processes in QCD, but with the special addition of
contributions from the direct-direct and direct-resolved $\gg$ processes.

\vskip1cm\leftline{{\bf Appendix II.
The Photon-Hadron Coherence Length}}
\vskip.5cm

\REF\Delduca{%
V. Del Duca, S. J. Brodsky, and P. Hoyer,
\sl Phys. Rev. \bf D46 \rm (1992) 931. }

At very high energies  the hadronic component of a photon state
resembles a coherent sum of vector
mesons.  The coherence time, as discussed by Ioffe\refmark\Ioffe~
and by Yennie \etal\refmark\Yennie, is
$
\Delta\tau = {1\over \Delta E} = {2P^+_\gamma\over Q^2 + \M^2}\ $
for intermediate  vector states of mass $\M.$  Thus in high energies
photon-nucleus reactions, a real or virtual
photon will generally convert to a hadronic system well before
interacting in the target, and the energy and nuclear size dependence
of  the photon-induced cross sections will resemble that of meson-induced
reactions. In fact,
as shown in Ref. [\Delduca], the coherence time
of a virtual photon  depends on whether its polarization is
longitudinal or transverse: $\tau_L = \tau_T/\sqrt 3.$  Thus
shadowing of the nuclear photoabsorption cross section will be
delayed to higher energies in the case of longitudinal current--nuclear
interactions.

The long coherence length between photons and the intermediate
vector states  at high energies and the
resulting photon-hadron duality can be
used as a general guide to the hadronic
interactions of photons at low transverse
momentum. In particular the long coherence length implies pomeron
factorization of photon-induced cross sections\refmark\BKT: $
\sigma_{\gamma\gamma} = {\sigma^2_{\gamma p}\over \sigma_{pp}} $
Thus one should be able to track the slow increase of the total
inelastic photon-photon cross section with that of the $\gamma p$
and $pp$ cross section.
\bigskip

\vskip1cm\leftline{{\bf 20. Figure Captions}}
\vskip.5cm

\noindent
{\bf Figure \figDD:}\brk
Factorization of the resolved photon-photon
amplitudes using the light-cone Fock
basis. (See Appendix I.)
In the case of the direct contributions, the photon annihilates within
the hard scattering amplitude.

\noindent
{\bf Figure \Afiga:}\brk
The $\gamma \gamma$ luminosity in Compton
             back--scattering of laser light; unpolarized
$e^\pm$ beams and laser photons (dashed), opposite helicities
of $e^\pm$ and $\gamma$ (full curve).
See Refs. [\two], [\seven], [\KMS].

\noindent
{\bf Figure \figH:}\brk
(a) Degree of circular polarization of the high
energy photons in polarized Compton back--scattering of laser
light for different $e^\pm$ and $\gamma$ helicity modes.
(b) Left/right asymmetry of the final state photon beam in
Compton back--scattering of laser light.
(c) Spectrum and degree of linear
polarization of the high-energy photons in
Compton back--scattering of linearly
polarized laser light; Ref. [\four].

\noindent
{\bf Figure \figBB:}\brk
Illustration of direct, resolved,
and higher-order loop contributions to high
energy $\gg$ collisions.

\noindent
{\bf Figure \figAA:}\brk
Representative cross sections for $W^+ W^-$production
and other electroweak reactions at a $\gg$ and $e^+ e^-$ linear collider.
The top mass is taken as $130$ GeV. The other subscripts refer to the
mass of the Higgs (in GeV). The Higgs mass is set to zero
for the reactions $e^+ e^- \to W^+W^- \nu \bar \nu$ and
$e^+ e^- \to Z Z \nu \bar \nu.$
{}From Refs. [\JikiaWWWW,\BB,\BaillargeonRev].

\noindent
{\bf Figure \figY:}\brk
Differential cross sections for producing a $W$ pair of a specific
helicity
combination at $\sqrt s_\gg= 500$ GeV as a function of
$\cos\theta$.
The curves are:\brk
1: (++++)+(----), 2: (+++-)+(++-+)+(--+-)+(---+),\brk
3: (++--)+(--++), 4: (+-++)+(-+++)+(+---)+(-+--),\brk
5: (+-+-)+(-+-+)+(-++-)+(+--+),\brk
6: (+-+-)+(-+0-)+(+--0)+(-++0)+(+-+0)+(-+-0)+(+-0-)+(-+0+),\brk
7: (+-00)+(-+00), 8: (++0+)+(--0-)+(+++0)+(---0),\brk
9: (++00)+(--00), 10: (++0-)+(++-0)+(--0+)+(--+0).\brk
The notation indicates $(\lambda_1\lambda_2\lambda_3\lambda_4)$,
where $\lambda_1, \lambda_2, \lambda_3$ and $\lambda_4$ are the
helicities
of the two photons and the $W^+$ and the $W^-$ respectively.
{}From Ref.  [\Yehudai].~

\noindent
{\bf Figure \figCC:}\brk
Standard Model one-loop contributions to the reaction $\gg \to Z Z$
including ghost $c^\pm$ and scalar $w$ contributions in the
background nonlinear gauge. From Ref.  [\JikiaNG].

\noindent
{\bf Figure \figEE:}\brk
The effective cross section for
$\gamma\gamma\rarrow Z^0 \gamma $  at an NLC taking into account
the back-scattered laser spectrum. The fermion and $W$ loop
contributions are shown for the production of a transversely (T) and
longitudinally (L) polarized $Z^0.$  The incident photons are taken to
have  positive helicity. From Ref.  [\JikiaNG].

\noindent
{\bf Figure \Afigd:}\brk
Number of events per year for the Standard Model
Higgs boson $(\Phi^0 \rightarrow b \overline{b}, t \overline{t}, ZZ)$
and for the heavy--quark backgrounds; Ref. [\three].
Here ${\cal L}_{\rm eff}=20\, fb^{-1}$,
$z_0=0.85$, $\VEV{\lambda_1\lambda_2} = 0.8$,
$\Gamma_{\rm expt} = 5\ GeV$.

\noindent
{\bf Figure \Afige:}\brk
Expected event rates for the Higgs signal
and the background processes in $b \overline{b}, c \overline{c}$
two--jet final states for polarized $\gamma$ beams;
Ref. [\fourteen].

\noindent
{\bf Figure \Afigf:}\brk
Invariant mass distribution in $\gamma \gamma
\rightarrow H \rightarrow ZZ$ and in the continuum  $ \gamma
\gamma \rightarrow ZZ$ for transverse and longitudinal $Z$
polarization; Ref. [\JikiaNG].

\noindent
{\bf Figure \Afigg:}\brk
The $\gamma \gamma$ polarization asymmetry ${\cal A}$
in Compton back--scattering of linearly polarized laser light
for various values of $x_0$; Ref. [\four].

\noindent
{\bf Figure \Afigh:}\brk
${\cal MSSM}$ Higgs particle $h^0$: signal and
background cross sections for $ b \overline{b}$ final states (a),
and the polarization asymmetry ${\cal A} (h^0)$ including the
background process (b); Ref. [\four].

\noindent
{\bf Figure \Afigi:}\brk
${\cal MSSM}$ Higgs particle $A^0$: signal and
background cross sections for $ b \overline{b}$ final states (a),
and the polarization asymmetry ${\cal A} (A^0)$ including the
background process (b); Ref. [\four].

\noindent
{\bf Figure \figEEE:}\brk
Illustration of $W W$ scattering at a photon-photon collider.
The kinematics of the
interacting $W$'s can be determined by tagging the spectator $W$'s.
The interacting pair can
scatter or annihilate, for example into a Higgs boson.

\noindent
{\bf Figure \figA:}\brk
(a) Deep-inelastic electron-photon scattering $e \gamma \to e X$. \brk
(b) The charged current
process $e \gamma \to \nu X$ in deep--inelastic
$e \gamma$ scattering.

\noindent
{\bf Figure \figB:}\brk
Event plane ${\cal P} = \left[\log Q^2,\log(1/x-1)\right]$ in
$e\gamma$ scattering ($Q^2$ is defined in GeV$^2$). Shown are
the two parallelograms which can be explored at LEP200 and LC500
and within which perturbative QCD can be applied.

\noindent
{\bf Figure \figC:}\brk
QCD prediction for the photon structure $F_2^\gamma(x,Q^2)$ at
$Q^2 = 200$~GeV$^2$ and sensitivity to the QCD $\Lambda$
parameter. Error bars correspond to an integrated luminosity of
500~pb$^{-1}$ at LEP200 and the range $100 < Q^2 < 500$~GeV$^2$.
{}From Ref. [\Cordier].

\noindent
{\bf Figure \figD:}\brk
Comparison of the $Q^2$ evolution of the photon structure
function in QCD with a model in which the coupling constant is
frozen.

\noindent
{\bf Figure \figE:}\brk
Experimentally observed $Q^2$ evolution of the photon structure
function; from Ref. [\sasaki].

\noindent
{\bf Figure \figI:}\brk
Comparison of  perturbative QCD predictions
with PLUTO data for the photon structure function at $Q^2=5.9$ GeV$^2.$
The charm quark contribution from leading and higher order
QCD  is also shown.  From Ref.  [\laenen].

\noindent
{\bf Figure \figF:}\brk
Theoretical estimate of the sensitivity to the effective
QCD scale
parameter, (a) from the evolution of $F_2^\gamma$ at large $x$;
(b) from the absolute size of $F_2^\gamma$ if the hadronic
component is assumed to be uncertain within $\pm 50\%$ at
$Q^2 = 100$~GeV$^2$.

\noindent
{\bf Figure \figL:}\brk
QCD contributions to jet transverse momentum
cross section in $\gg$ collisions.
The resolved contributions are based on
the Drees-Godpole model for the gluon distribution in the photon. From
Ref. [\Resolved].

\noindent
{\bf Figure \figO:}\brk
The effect of multiple scattering on the
mini-jet contribution to the $\gg$
total cross section. The jet contributions are shown for various
$p_T$ minimum cut-offs, with (solid line) and without
(dashed line) the effect of eikonalization. From
Ref.  [\FS].

\noindent
{\bf Figure \figWW:}\brk
Perturbative QCD contributions to large momentum transfer
exclusive double diffractive $\gg$ processes. The two-gluon
exchange pomeron  contributions to vector meson pair production and
three-gluon exchange odderon contributions to neutral pseudoscalar and
tensor meson pair production are illustrated.

\noindent
{\bf Figure \figR:}\brk
Direct and resolved contributions to heavy quark production in $\gg$
collisions.

\noindent
{\bf Figure \figS:}\brk
QCD leading and next-to-leading order contributions
to the inclusive charm
production cross section. The resolved contributions are based on
the Drees-Godpole model for the gluon distribution in the photon. From
Ref. [\DKZZ].

\noindent
{\bf Figure \figT:}\brk
TPC$\gg,$  TASSO, and TRISTAN data for $\sigma(\epem\rarrow \epem
D^{\ast\pm}X) $ compared with the QCD prediction of Drees \etal\
{}From Ref.  [\DKZZ].
The dashed lines show the once-resolved contributions.
The upper line is $\mu = m_c = 1.3$ GeV.  The lower line is
$\mu = 2m_c,\ m_c = 1.8$ GeV.

\noindent
{\bf Figure \figU:}\brk
(a) Effective cross section $\VEV{\sigma(\gg \to t {\bar t})}/
\sigma(\epem)_{pt}$
with (solid) and without (dashed) QCD corrections for $m_t=150~GeV.$ The
convolution with a back-scattered laser spectrum $(\omega=1.26$ eV) is
included.
(b) The  effective differential cross section
${d\VEV{\sigma(\gg \to t {\bar t})}/dz / \sigma(\epem)_{pt}}$
and the resonance
signal
predicted at $\sqrt{s_\epem} = 500$ GeV. Here $z = \M_{t \bar
t}/\sqrt{s_\epem}.$  From Ref.  [\KMS].

\noindent
{\bf Figure \figQQ:}\brk
The single top production cross section
and its competing
backgrounds in high energy electron-photon collisions.
{}From Ref. [\Cheung].
\bigskip

\refout
\end